\documentclass[10pt,aps,amsmath,amssymb,nccmth,prb,twocolumn,notitlepage,showpacs,superscriptaddress]{revtex4-2}
\usepackage{amsmath, nccmath}
\usepackage{amssymb}
\usepackage{graphicx}
\usepackage{epstopdf}
\usepackage[colorlinks=true]{hyperref}
\usepackage{physics}
\usepackage{mathrsfs}
\usepackage{comment}
\usepackage{float}
\usepackage{placeins}
\usepackage[percent]{overpic}
\usepackage{subcaption}
\DeclareCaptionJustification{revtexjust}{\leftskip=0pt \rightskip=0pt \parfillskip=0pt plus 1fil\relax}
\renewcommand\({\begin{equation}}	
\renewcommand\){\end{equation}}
\renewcommand\[{\begin{eqnarray}}	
\renewcommand\]{\end{eqnarray}}
\newcommand{\al}[1]{\begin{aligned}#1\end{aligned}}
\usepackage[dvipsnames]{xcolor}

\newcommand{\reapp}[1]{{\color{black}{#1}}}

\begin{document}

\title{Robust Tripartite Entanglement Generation via Correlated Noise in Spin Qubits}

\author{Sander Driessen}
\affiliation{Department of Physics, University of Basel, Klingelbergstrasse 82, 4056 Basel, Switzerland}
\affiliation{Department of Applied Physics, Eindhoven University of Technology, Eindhoven 5612 AZ, The Netherlands}
\author{Ji Zou}
\email{ji.zou@kfupm.edu.sa}
\affiliation{Physics Department, King Fahd University of Petroleum and Minerals, 31261, Dhahran, Saudi Arabia}
\affiliation{Quantum Center, KFUPM, Dhahran, Saudi Arabia}
\author{Even Thingstad}
\affiliation{Department of Physics, University of Basel, Klingelbergstrasse 82, 4056 Basel, Switzerland}
\author{Jelena Klinovaja}
\affiliation{Department of Physics, University of Basel, Klingelbergstrasse 82, 4056 Basel, Switzerland}
\author{Daniel Loss} \affiliation{Department of Physics, University of Basel, Klingelbergstrasse 82, 4056 Basel, Switzerland}
\affiliation{Physics Department, King Fahd University of Petroleum and Minerals, 31261, Dhahran, Saudi Arabia}
\affiliation{Quantum Center, KFUPM, Dhahran, Saudi Arabia}
\affiliation{RDIA Chair in Quantum Computing}

\begin{abstract}
We investigate the generation of genuine tripartite entanglement in a triangular spin-qubit system due to spatially correlated noise. In particular,  we demonstrate how the formation of a highly entangled dark state, a $W$ state, enables robust, long-lived tripartite entanglement. Surprisingly, we find that environmentally induced coherent coupling does not play a crucial role in sustaining this entanglement. This contrasts sharply with the two-qubit case, where the induced coupling significantly influences the entanglement dynamics. Furthermore, we explore two promising approaches to enhance the tripartite entanglement by steering the system towards the dark state: post-selection and coherent driving. Our findings offer a robust method for generating high-fidelity tripartite entangled states with potential applications in quantum computation.
\end{abstract}

\date{\today}
\maketitle

\section{Introduction}Quantum computing promises to solve problems intractable for classical computers by harnessing entanglement~\cite{Preskill2018QuantumBeyond,Shor1994AlgorithmsFactoring, Altman2021QuantumOpportunities,Daley2022PracticalSimulation}. Among the various platforms~\cite{blinov2004observation,blatt2008entangled,koch2007charge,barends2013coherent,zou2026perspective,psaroudaki2021skyrmion,qu2024density,zou2023quantum}, spin qubits stand out due to their long coherence times, scalability, and compatibility with existing semiconductor technology~\cite{Loss1998QuantumDots,Stano2022ReviewNanostructures,Burkard2023SemiconductorQubits,bosco2024high,zou2024topological,PRXQuantum.4.020305}. However, qubits inevitably interact with noisy environments, causing errors that require quantum error correction (QEC)~\cite{Schlosshauer2019QuantumDecoherence}. While QEC can mitigate local uncorrelated noise, spatially correlated noise remains particularly challenging for existing schemes~\cite{Laflamme1996PerfectCode, Clemens2004, Klesse2005QuantumNoise, Aharonov2006Fault-tolerantNoise, Preskill2013SufficientComputing}. Such pairwise correlated noise has been observed in semiconductor spin qubits \cite{rojas2025inferring,Yoneda2023Noise-correlationSilicon, Rojas-Arias2023SpatialQubits}, and possible ways to mitigate it are under active exploration~\cite{Harper2023LearningProcessor,Seif2024SuppressingCompiling}.  
Interestingly, insights from quantum optics suggest that properly engineered correlated noise can serve as a resource for entanglement generation~\cite{Poyatos1996QuantumIons, Plenio1999Cavity-loss-inducedAtoms, Diehl2008, Kraus2008, Verstraete2009}. 
Recent studies in spin qubit systems also demonstrate the dissipative generation of Bell states via spatially correlated noise~\cite{Zou2024SpatiallyQubits,Zou2022Bell-stateCoupling}. 
However, this long-lived bipartite entanglement is highly sensitive to inversion symmetry breaking in the environment, which induces Dzyaloshinskii–Moriya (DM) interactions that ultimately spoil the steady-state generation of Bell states. Beyond bipartite entanglement, scalable quantum technologies often require \textit{multipartite} entanglement~\cite{Cirac1999DistributedChannels, Cavalcanti2011QuantumNonlocality, Aolita2012FullyCorrelations}, nonclassical correlations shared among multiple qubits~\cite{Horodecki2009QuantumEntanglement, Ma2024MultipartiteReview}, which enables  quantum secret sharing \cite{Lipinska2018AnonymousState, Miguel-Ramiro2020DelocalizedNetworks}, error-resistant protocols \cite{Liao2022BenchmarkingProtocols}, and various quantum tasks \cite{Agrawal2006PerfectStates, Choi2010EntanglementMemories,Ng2014QuantumW-state}. 

This raises an important question: can pairwise correlated noise be leveraged to generate genuine multipartite entanglement? Furthermore, since the multipartite entanglement is of a fundamentally different nature than the bipartite entanglement in a two-qubit system~\cite{Verstraete2002,bengtsson2006chapter15, Rothlisberger2009,Plenio2014}, it is not clear whether an environmentally induced DM interaction is detrimental also to the generation of multipartite entanglement through correlated noise.  Beyond the fundamental significance, understanding how multipartite entanglement emerges and persists in realistic noisy environments is essential for the development of scalable and fault-tolerant quantum technologies.

\begin{figure}
    \centering
    \includegraphics[width=\linewidth]{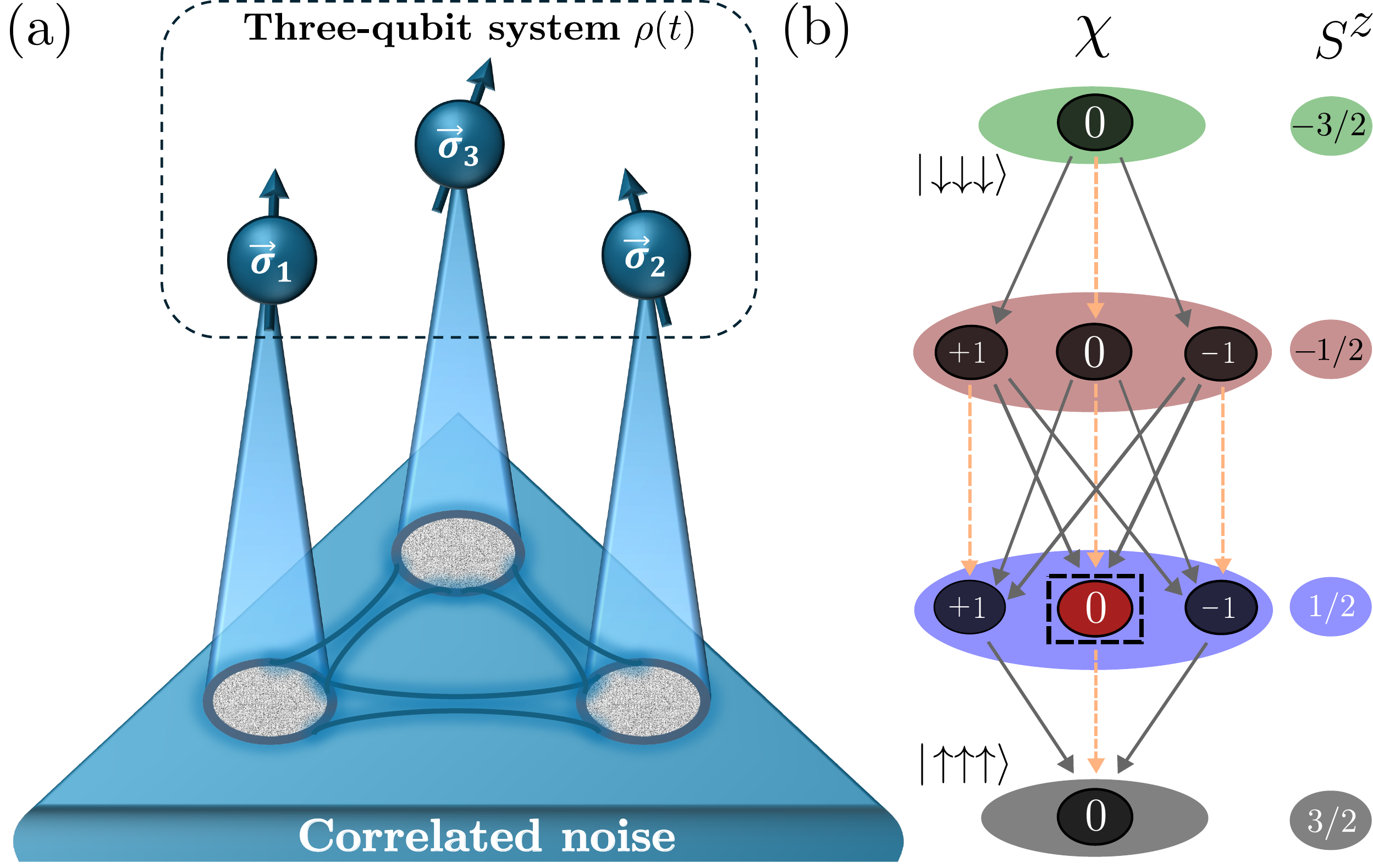}
    \caption{ (a) A system of three spin qubits is arranged in an equilateral triangle and coupled to a noisy environment. (b) The  states of the three-qubit system [spheres in (b)] can be labeled by the eigenvalues $\chi \in \{0, \pm 1\}$ of the chirality operator $\hat\chi =\vec{\sigma}_1 \cdot \vec{\sigma}_2 \times \vec{\sigma}_3/2\sqrt{3}$ and the eigenvalues $S^z \in \{\pm 3/2, \pm 1/2\}$ of the total spin-$z$ operator $\hat{S}^z = \sum_i \hat{\sigma}_i^z/2$. Noise induces transitions (arrows) between the eigenstates, while correlated spatial noise introduces an imbalance in the transition amplitudes. When the non-local dissipation is strong, one type of transition can be suppressed (orange dashed arrows), and this allows the formation of a dark state (dashed box). }
    \label{fig:1}
\end{figure}

In this work, we demonstrate that pairwise spatially correlated noise, arising from a generic noisy medium, can drive multipartite entanglement generation in a quantum system of three spin qubits, as depicted in Fig.~\ref{fig:1}(a). 
{This setup provides the simplest experimentally realizable~\cite{Seo2013,Barthelemy2013,Flentje2017,russ2017three, Noiri2017,Acuna2024,hsieh2012physics} geometry where the phase of the correlated noise becomes relevant, which is essential for our proposal.  } 
By microscopically deriving a full master equation governing the qubit system evolution~\cite{Breuer2007TheSystems,Campaioli2024QuantumBeyond}, we provide a unified framework that accounts for both local and spatially correlated noise. Surprisingly, we find that, in stark contrast to Bell-state generation, long-lived genuine tripartite entanglement emerges without precise control over coherent interactions and is immune to the DM interaction. This entanglement generation is robust and occurs irrespective of the initial state. Furthermore, we show how to distill a near-perfect $W$ state with fidelity exceeding 99\% via post-selection or driving. 
{While the proposal is general, our concrete near-term target platform is  spin qubits in gate-defined quantum dots. }

\section{Model}We consider  a model of three {driven} spin qubits arranged in an equilateral triangle, weakly coupled to a noisy environment, as shown in Fig.~\ref{fig:1}(a).  
{For spin qubits in gate-defined quantum dots, the dominant qubit-environment coupling is longitudinal. As shown in Appendix~\ref{sec:master_eq}, however, a coherent transverse drive maps the lab-frame Hamiltonian onto an effective transverse coupling form within the rotating wave approximation, with the qubit splitting replaced by the Rabi frequency. Therefore, we consider total Hamiltonian to be $H = H_S + H_E + H_{\text{SE}},$}
where the qubits are governed by $H_S = -\Delta\sum_{i=1}^3\sigma_i^z/2$ with energy splitting $\Delta >0$. Here, $H_E$ is an unspecified Hamiltonian describing the noisy environment, while the system-environment coupling is described by
 $  H_{\text{SE}}=\lambda\sum_{i=1}^3 (\sigma_i^+\otimes E_i^- + \sigma_i^-\otimes E_i^+),$ where $\lambda$ is the coupling strength, $\sigma_i^\pm$ are the spin ladder operators for qubit $i$, and $E_i^\pm$ are Hermitian conjugate local operators in the environment.
We further assume the spin qubits are separated by hundreds of nanometers, so both the direct exchange interaction and the dipolar coupling are negligible.

We microscopically derive the Lindblad master equation for the qubit density matrix $[\rho_{ij}]_{8\times 8}$  under the Born-Markov approximation (see Appendix~\ref{sec:master_eq}), yielding
 $d\rho/dt = -(i/\hbar)[H_S + H_{\mathrm{eff}},\rho] + \mathcal{L}[\rho].$
Here, $H_{\mathrm{eff}}=\sum_{i}[\mathcal{J}_\perp(\sigma_i^x\sigma_{i+1}^x + \sigma_i^y\sigma_{i+1}^y) + D\hat{z}\cdot \Vec{\sigma}_i\times \Vec{\sigma}_{i+1}]/2$ describes the effective coherent coupling induced by the noisy environment, comprising an in-plane XX interaction and a DM interaction between all three qubits.  This coherent coupling can be recast in the form
\begin{equation}
    H_{\mathrm{eff}}= \mathcal{J}\sum_{i=1}^3(e^{i\psi}\sigma_i^+\sigma^-_{i+1}+e^{-i\psi}\sigma_i^-\sigma^+_{i+1}),
\end{equation} 
where $\mathcal{J} = \sqrt{\mathcal{J}_\perp^2 + D^2 }$ denotes the coherent coupling amplitude and $\psi = \operatorname{arg}(\mathcal{J}_\perp + iD)$ denotes the corresponding phase. Explicit expressions for these couplings in terms of environmental correlation functions are given in Appendix~\ref{sec:master_eq}.
The environment also leads to qubit {relaxation} through the dissipator
\begin{equation}
  \al{  \mathcal{L}[\rho] = & \sum_{ij}\gamma_{ij}\big(\mathcal{O}_j\rho\mathcal{O}_i^\dag-\frac{1}{2}\{\mathcal{O}_i^\dag \mathcal{O}_j,\rho\}\big) \\
      &+  \sum_{ij}\tilde\gamma_{ij}\big(\tilde{\mathcal{O}}_j\rho\tilde{\mathcal{O}}_i^\dag-\frac{1}{2}\{\tilde{\mathcal{O}}_i^\dag \tilde{\mathcal{O}}_j,\rho\}\big), }
    \label{eq:dissipator}
\end{equation}
where $\mathcal{O}_i\in\{\sigma_1^+,\sigma_2^+,\sigma_3^+\}$  and $\tilde{\mathcal{O}}_i\in\{\sigma_1^-,\sigma_2^-,\sigma_3^-\} $ are qubit operators. The first term describes qubit relaxation (i.e. transitions to states with lower energy, as determined from $H_S$), while the second term corresponds to the reverse excitation processes. Importantly, the elements $\gamma_{ij}$ in the dissipator are directly related to the positive frequency component of the noise power spectral density (PSD) $S_{ij}(\Delta)$ of the environment at the qubit transition frequencies, as given by  \cite{Clerk2010IntroductionAmplification,Zou2024SpatiallyQubits}
\begin{equation}
    \gamma_{ij} =\frac{\lambda^2}{\hbar^2} S_{ij}(\Delta) = \frac{\lambda^2}{\hbar^2} \int_{-\infty}^{\infty}\! dt\, e^{i\Delta t}\langle E^+_i(t)E^-_j\rangle.
\end{equation} 
In contrast, $\tilde{\gamma}_{ij}$ is linked to the energy absorption rate, which is related to $\gamma_{ij}$ via $\tilde{\gamma}_{ij} = \gamma_{ji} \exp(-\beta\hbar\Delta)$ in accordance with the fluctuation-dissipation theorem.
Thus, only decay processes are relevant at zero temperature, $\beta\equiv 1/(k_BT)\rightarrow \infty$. 
The auto-PSD corresponding to the diagonal terms ($i=j$) 
affect each qubit individually. In contrast, the cross-PSD associated with the off-diagonal terms ($i\neq j$)
captures correlated noise between qubits $i$ and $j$, and arises from shared environmental modes. We remark that the effect of these off-diagonal elements is also being actively explored in the large-spin limit, where they give rise to non-reciprocal phenomena through dissipative coupling between macrospins~\cite{Zou2024Dissipative,YU20241,PhysRevResearch.6.033207,Yuan2023Unidirectional,zou2025emergent}.

For simplicity, we assume a homogeneous environment in the following, so that the dissipator matrix 
takes the form
\begin{equation}
    [\gamma_{ij}] = 
    \begin{pmatrix}
        a & A & A^* \\
        A^* & a & A \\
        A & A^* & a
    \end{pmatrix},
\end{equation}
where we denote the decay rates induced by local and spatially correlated noise by $a$ and $A$. In Appendix~\ref{sec:non-homogeneous}, we show that even when we introduce a small asymmetry in the model, our results remain robust. We further note that since $S_{ij}(\Delta) = S_{ji}^*(\Delta)$, $a$ is real-valued, while $A \equiv |A|e^{i\phi}$, can take on complex values, as also observed experimentally~\cite{VonLupke2020Two-QubitQubits}. For the dynamics described by the master equation to remain physically valid, the evolution must be completely positive \cite{Breuer2007TheSystems}. This imposes the requirement  $a\geq 2|A|$.

To elucidate the zero-temperature {relaxation} dynamics, we diagonalize the dissipator 
matrix $[\gamma_{ij}]$ by introducing the jump operators
\begin{equation}
    J_k = \sqrt{\frac{\gamma_k(a,A)}{3}}(\eta^k\sigma_1^+ + \eta^{-k}\sigma_2^+ + \sigma_3^+),
    \label{eq:J_k}
\end{equation}
where  $k\in \{-1,0,1\}$, $\eta = \exp(2\pi i/3)$, and the amplitudes $\gamma_k(a, A)$ are given by the eigenvalues of $[\gamma_{ij}]$, 
\begin{equation}
    \gamma_k(a,A) =  {a + 2|A|\cos{(\phi +2\pi k / 3)}} .
    \label{eq:gamma_k}
\end{equation}
{We emphasize that,
like an Aharonov--Bohm flux, the phase $\phi$ is a gauge invariant quantity associated with a closed triangular loop
and cannot be removed by any local change of qubit basis. This nonremovable loop phase is the essential ingredient that allows one of the eigenvalues $\gamma_k$ to vanish at $|A|=a/2$ and $\phi=\pi$ (or $\phi=\pm\pi/3$), thus producing the dark $W$ state.}
The dissipator then takes the form 
$ \mathcal{L}[\rho] = \sum_{k} J_k\rho J_k^\dag-(1/2)\{J_k^\dag J_k,\rho\}$,
where the first term describes incoherent quantum jumps induced by the environment, while the second term accounts for the associated damping, ensuring conservation of probability. At finite temperature, excitations become relevant. The corresponding jump operators follow from the diagonalization of the terms in the dissipator which describe excitations [see Eq.~\eqref{eq:dissipator}], and are given by the Hermitian conjugate of the jump operators in Eq.~\eqref{eq:J_k} upon the additional replacement $\gamma_k \rightarrow \tilde{\gamma}_{k} = \gamma_k \exp(-\beta \hbar \Delta)$.

The form of the jump operators suggests that it is useful to describe the system in terms of chiral basis states. To this end, we introduce the total spin-$z$ operator (in units of $\hbar$) and the scalar spin chirality operator, respectively,
\begin{equation}
    \hat{S}^z= \frac{1}{2}\sum_i \sigma^z_i  
    \quad \text{and} \quad
    \hat{\chi}= \frac{1}{2\sqrt{3}}\Vec{\sigma}_1\cdot(\Vec{\sigma}_2\times\Vec{\sigma}_3).
\end{equation} 
These commuting operators have joint eigenstates $\ket{S^z, \chi}$, labeled by $S^z \in \{\pm 1/2, \pm 3/2\}$ and integers $\chi$ satisfying $|\chi| \leq 3/2 - |S^z|$, which form a complete basis of the three-qubit Hilbert space, as illustrated in Fig.~\ref{fig:1}(b).
Importantly, the operators $\hat{S}^z$ and $\hat{\chi}$ also commute with $H_S + H_\mathrm{eff}$, so that the basis states $|S^z, \chi \rangle$ are also energy eigenstates when the dissipation is negligible. 
The states are depicted in the energy diagram in Fig. \ref{fig:1}(b), where the energetic ground state is $\ket{\uparrow\uparrow\uparrow}$  and the state with highest energy is $\ket{\downarrow\downarrow\downarrow}$.
The remaining states are $W$ states of the form 
\begin{align}
    \ket{S^z=\tfrac{1}{2},\chi} = \frac{1}{\sqrt{3}} \left({\eta^{-\chi}\ket{\uparrow\uparrow\downarrow} + \eta^{\chi}\ket{\uparrow\downarrow\uparrow} + \ket{\downarrow\uparrow\uparrow}} \right),
\label{eq:eigstate_Sz+} 
\end{align}
while the corresponding basis states with $S^z = -1/2$ are obtained by flipping all spins in the above expression.

Acting on the chirality and total spin-$z$ states, the jump operators assume the particularly simple form 
 $ J_k\!\ket{S^z\!,\! \chi }\! \propto\! \ket{ S^z+1,\! \chi+k }, $
where $\chi + k$ is taken
modulo 3~\footnote{Example: If $\chi=1$ and $J_1$ is applied, then $1+1$ cycles to $-1$ rather than $2$.}. 
A picture now emerges where the environment-induced {relaxation} dynamics manifests as transitions between states with different chirality and total spin-$z$. These decay processes are indicated in Fig.~\ref{fig:1}(b) for a particular set of parameters such that $\gamma_0 = 0$. 

\begin{figure}
    \centering
    \includegraphics[width=1\linewidth]{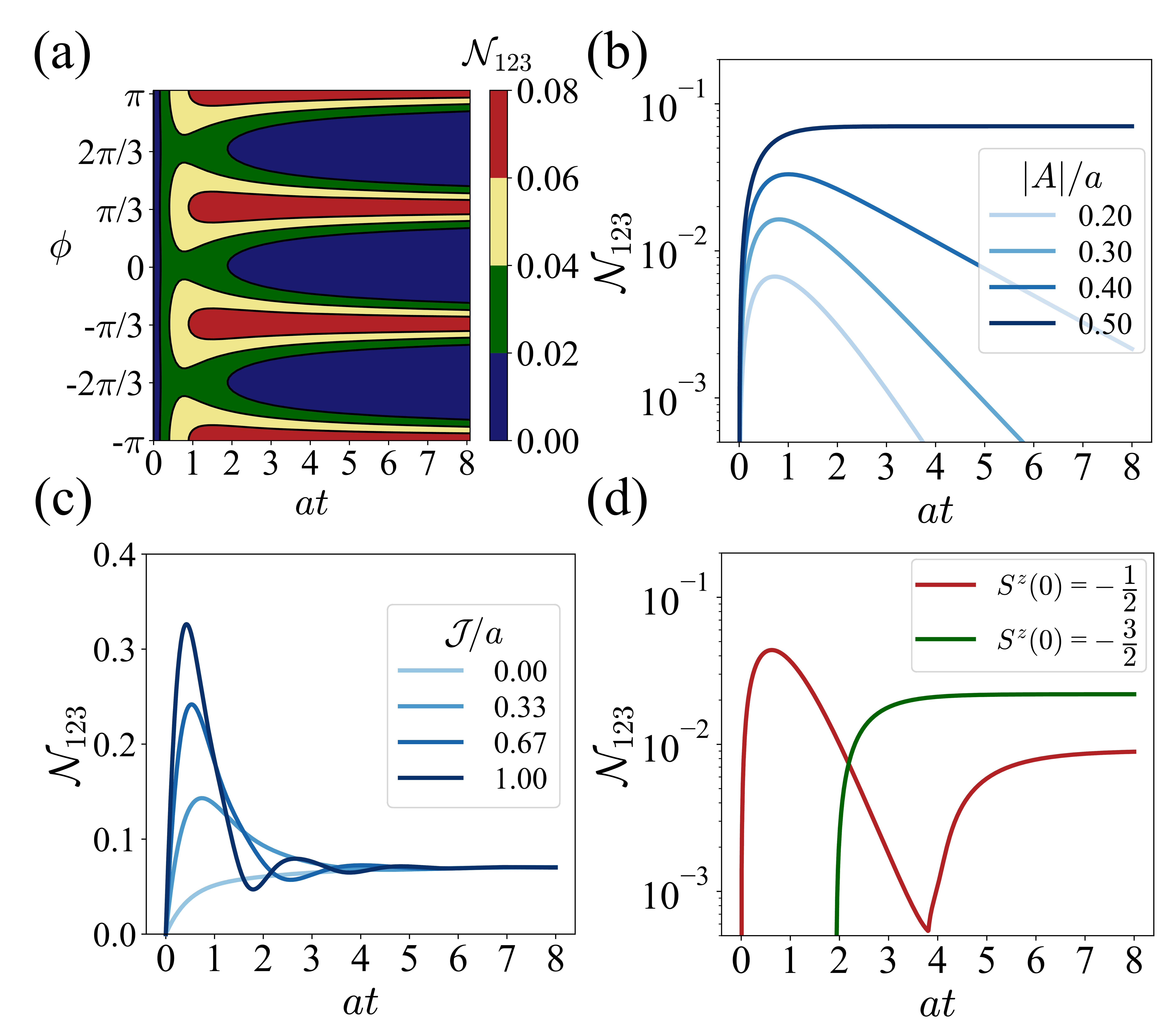}
    \caption{ Entanglement dynamics demonstrating long-lived entanglement for parameters such that one of the jump amplitudes  $ \gamma_k$ in Eq.~\eqref{eq:gamma_k} vanishes. (a) Evolution of the tripartite negativity $\mathcal{N}_{123}$ for different values of $\phi$  when the system is initialized in the state $\ket{\downarrow\uparrow\uparrow}$. (b) Entanglement dynamics for varying correlated noise. (c) Entanglement dynamics for varying coherent interaction amplitude $\mathcal{J}$. (d) Upon initialization in a spin product state in the total spin-$z$ sectors $S^z=-1/2$ (red) and $S^z=-3/2$ (green), the system exhibits revival and bursts of entanglement. Parameters used throughout unless stated otherwise: $\mathcal{J} = 0$, $|A|/a=0.5$, $\phi = \pi$, and $\Delta/a= 100 $. 
    }
    \label{fig:2}
\end{figure}

{Finally, we stress that our model is distinct from the standard open Dicke and homogeneous central-spin
models~\cite{schneider2002entanglement,kessler2012dissipative}. Those models have $SU(2)$ structure where the collective jump operator $\sum_i \sigma_i^-$ commutes with the total-spin operator $\hat S^2 \equiv \hat S_x^2 + \hat S_y^2 + \hat S_z^2$. Thus the Hilbert space splits into three completely decoupled total-spin sectors. In contrast, our model has a $U(1)$ structure, where the jump operators do not commute with $\hat S^2$ 
and the spin sectors do not decouple. 
Moreover, the dark state in our model arises because one of the eigenvalues $\gamma_k$ of the dissipator matrix vanishes due to a finite gauge-invariant loop phase $\phi$. This has no analogue in the small-sample Dicke limit.}

\section{Entanglement dynamics with correlated noise}To quantify genuine tripartite entanglement we use the tripartite negativity given by the geometric average \cite{Sabin2008ASystems, Ma2024MultipartiteReview}
\begin{equation}
    \mathcal{N}_{123} = \sqrt[3]{\mathcal{N}_{[1]23}\mathcal{N}_{1[2]3}\mathcal{N}_{12[3]}}.
    \label{eq:tripartite_negativity}
\end{equation}
Here, the bipartite negativity $\mathcal{N}_{[j]kl}$ quantifies the entanglement of qubit $j$ with the other two    qubits~\cite{Vidal2002ComputableEntanglement}, and is defined by 
$\mathcal{N}_{[j]kl} = 2\sum_i |\text{min}\{0,\lambda^{T_j}_i\}|$,
where $T_j$ denotes the partial transpose with respect to qubit $j$ and $\lambda^{T_j}_i$ is the $i$th eigenvalue of $\rho^{T_j}$.

In the following, we solve the dynamics of the three-qubit system and its tripartite entanglement analytically at zero temperature, where only decay processes survive.  We also simulate the finite-temperature dynamics numerically using QuTip~\cite{Lambert2024QuTiPPython} to confirm  that the results are robust to temperatures below the qubit splitting $\Delta$, as discussed in further detail in Appendix~\ref{sec:finite_temp}.
We initialize the system in an unentangled spin product state, and first consider $\ket{\uparrow\uparrow\downarrow}$ for concreteness. 
This state can be written as an equal superposition of the chirality basis states in the subspace defined by $S^z = 1/2$.
In the absence of spatially correlated noise, the jump amplitudes $\gamma_k$ in Eq.~\eqref{eq:gamma_k} are uniform. Thus, the three states $\ket{\tfrac{1}{2},\chi=0,\pm 1}$ comprising the superposition decay equally fast, and this prevents the development of any entanglement. In contrast, spatially correlated noise introduces an imbalance in the decay rates. Under time evolution, this distorts the equal superposition of the initial state, and eventually allows a single (or two) chiral basis states to dominate. Since the chiral basis states are $W$ states with tripartite negativity $\mathcal{N}_{123}^{\mathrm{W}} = 2\sqrt{2}/3\approx0.94$ \cite{Sabin2008ASystems}, a genuine tripartite entanglement develops under time evolution. 
The decay rate imbalance can change the {relaxation} dynamics significantly. 
In particular, when the non-local decay is $|A|=a/2$  and the phase of the noise is $\phi=\pm \pi/3$ or $\phi=\pi$, one of the jump amplitudes vanishes. This results in the formation of an entangled dark state, as indicated by a dashed box (i.e. the achiral $W$ state when $\phi=\pi$) in Fig.~\ref{fig:1}. 
{We point out that the required ratio $|A|/a = 1/2$ between correlated and local noise is experimentally achievable~\cite{rojas2025inferring,Yoneda2023Noise-correlationSilicon, Rojas-Arias2023SpatialQubits}.
At this optimal point, time evolution generates} long-lived tripartite entanglement, as shown for $\mathcal{J} = 0$ in Fig.~\ref{fig:2}(a). 
As shown in Fig.~\ref{fig:2}(b), finite tripartite entanglement is generated also for weaker correlated noise (i.e. smaller values of $|A|/a$), with a lifetime that scales as $1/(a - 2|A|)$. {We stress that only the relative strength of $a$ and $A$ is essential for achieving long-lived entanglement. For $1/f$-type noise, we expect that
both rates share the same frequency scaling and that the entanglement
dynamics therefore remains qualitatively similar even if the two decay rates depend on the effective qubit frequency $\Delta$, as shown in Appendix~\ref{sec:non-homogeneous}.}
{The long-time entanglement dynamics remains essentially unchanged also in the presence of pure dephasing and realistic qubit-frequency detuning, provided that the correlated dephasing is comparable to the local dephasing and that the detuning is smaller than the strength of the correlated noise (see Appendix~\ref{sec:non-homogeneous}).
We also remark that the long-lived entanglement persists even when the non-local dissipation on the links is slightly inhomogeneous (see Appendix~\ref{sec:non-homogeneous}).
} 

In addition to the dissipative decay,
environmentally induced coherent interactions $H_\mathrm{eff}$ between qubits typically also play a crucial role in the entanglement dynamics. In Fig.~\ref{fig:2}(c), we show how the entanglement generation is affected by a finite coherent coupling $\mathcal{J}$. Interestingly, we note that while the transient oscillations in tripartite entanglement depend on $\mathcal{J}$, the entanglement in the long-time limit remains unaffected. This occurs because the dark state remains an eigenstate of the environmentally induced coherent interaction $H_\mathrm{eff}$. Thus, 
the entanglement in the long-time limit is determined by the statistical weight of this state immediately after initialization. We stress that this behavior is fundamentally different from the two-qubit case~\cite{Zou2022Bell-stateCoupling}, where the induced DM interaction inevitably drives the system out of the dark state and causes entanglement loss.

Upon initializing the system in spin product states within the total spin-$z$ sectors  $S^z\!=\!-1/2$ or $-3/2$, we find intricate entanglement dynamics qualitatively different from the result obtained by initialization in the sector  $S^z\!\!=\!\!1/2$. Specifically, as shown in Fig.~\ref{fig:2}(d), we find bursts and revival of entanglement. When the system is initialized in the sector $S^z \!\!=\!\! -1/2$ (red curve), an imbalance in eigenstate decay rates within the sector induces an immediate entanglement increase. As this sector depletes, its contribution to tripartite entanglement diminishes, corresponding to the dip of the red curve. Meanwhile, the population transfer to the $S^z =1/2$ sector enables entanglement generation from the dark state, leading to a revival of entanglement. 
When the system is instead initialized in the sector $S^z = -3/2$ (green curve), there is no immediate development of entanglement, as the sector only has a single unentangled state. While the states within the sector $S^z = -1/2$ become unevenly populated, the subsequent decay is sufficiently fast to prevent accumulation of significant weight in this sector. Entanglement therefore first emerges due to development of population imbalance within the sector $S^z = 1/2$, and this gives rise to a sudden burst of entanglement.

While the correlated noise can generate long-lived tripartite entanglement regardless of initialization, the final value depends on the initial state. In particular, it depends on the fidelity $\mathcal{F}$ of the the dark state $|W\rangle$ in the long-time limit. 
Assuming that the density matrix approaches $\rho(t \rightarrow \infty) = \mathcal{F} |W \rangle \langle W | + (1 - \mathcal{F} )\ket{\uparrow \uparrow \uparrow } \bra{\uparrow\uparrow \uparrow}$ in the long time limit, the tripartite negativity is 
\begin{equation}
\mathcal{N}_{123}(\mathcal{F})= \sqrt{ ( \mathcal{N}_{123}^{\mathrm{W}}\mathcal{F})^2 + ({1-\mathcal{F}} )^2  }-  (1-\mathcal{F}).
\label{eq:N123_Fidelity}
\end{equation}
The $W$-state fidelity in the long-time limit for the different initializations can be understood based on the rates of decay between the states in Fig.~\ref{fig:1}(b). For instance, when the system is initialized in the $S^z = 1/2$ sector, there is a 1/3 probability of occupying the $W$ state, corresponding to an entanglement of $\mathcal{N}_{123} \approx 0.07$. Similarly, one may derive analytical expressions for the final entanglement when the system is initialized in the total spin-$z$ sectors $S^z=-1/2$ and $S^z=-3/2$, which yields $\mathcal{N}_{123}\approx0.009$ and $\mathcal{N}_{123}\approx0.02$, respectively. In general, the tripartite entanglement is clearly enhanced if the probability of ending up in the dark state increases. Below, we explore two schemes to achieve this.

\section{Enhancing tripartite entanglement}
\begin{figure}
    \centering
    \includegraphics[width=0.95\linewidth]{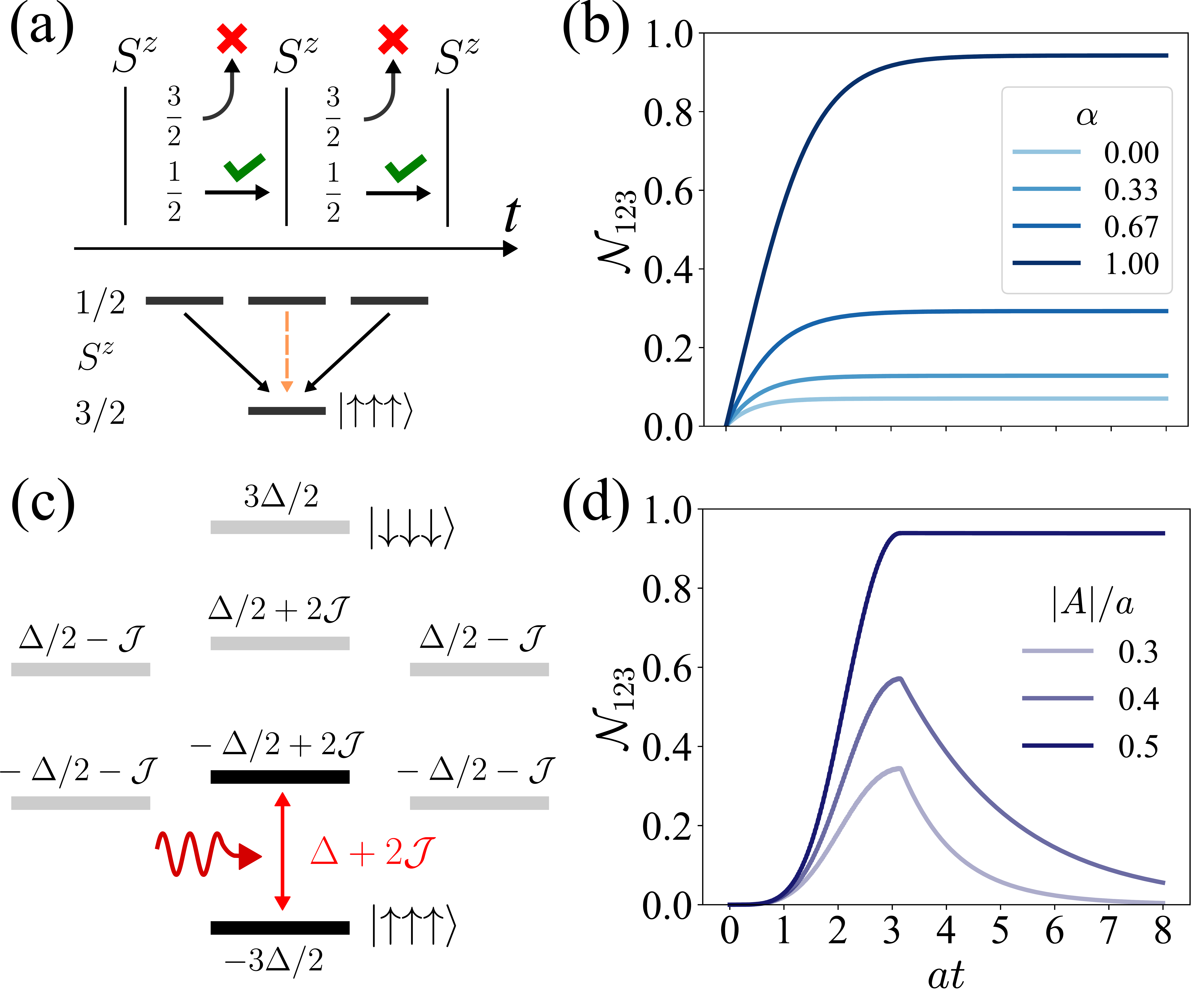}
    \caption{
    Enhancement of entanglement generation: (a)-(b) post selection and (c)-(d) driving.  (a) By periodically measuring the total spin-$z$ and post-selecting states with $S^z=1/2$, the decay from the spin-sector $S^z=1/2$ to the spin sector $S^z=3/2$ is inhibited. Since the decay rates are imbalanced, this effectively leads to transfer of probability to the state with the slowest decay rate. (b) Entanglement dynamics determined by Eq.~\eqref{eq:postSelectionEoM} for $|A|/a = 0.5$, and various $\alpha$. Post-selection increases the tripartite entanglement. (c) Energy level diagram in the presence of coherent coupling with $\psi = 0$. The states within a given total spin-$z$ sector are split. The system is driven with frequency corresponding to the splitting between the ground state and a single $W$ state. (d) Entanglement dynamics in the presence of driving for $\mathcal{J}/a =10$ and various values of the correlated noise. Parameters used throughout: $\phi= \pi$, $\mathcal{J}/a = 0$, and $\Delta/a = 100$ unless specified otherwise.}
    \label{fig:3}
\end{figure}

The fidelity of the dark state can be increased by utilizing measurement and control \cite{Wiseman2009QuantumControl, Zhang2019HeraldedWaveguide, Doggen2022,Doggen2023,Doggen2024}. As we have established, the decay rates fully control the entanglement generation. The decay rates can be reduced by measuring the total spin-$z$ component of the three-qubit system and post-selecting outcomes within a specific sector, as shown in Fig.~\ref{fig:3}(a). This process can be modeled by the equation \cite{Minganti2020Hybrid-liouvillianTrajectories, Liu2025LindbladianJumps}
\begin{equation}
    d\rho/dt\! =\! -(i/\hbar)[H_{\mathrm{non}},\rho] + (1-\alpha)\sum_{k}J_k\rho J^\dag_k,
    \label{eq:postSelectionEoM}
\end{equation}
where $H_{\mathrm{non}}\! =\! H_{S}+H_{\mathrm{eff}}-(i/2)\sum_kJ^\dag_k J_k$ is a non-Hermitian Hamiltonian and the corresponding commutator is defined as $[H_{\text{non}},\rho] = H_{\text{non}}\rho -\rho H_{\text{non}}^\dag$. The parameter $\alpha \in [0,1]$ {describes the probability of detecting a jump and quantifies the level of post-selection.} For $\alpha = 0$, the system evolves freely as given by the full Lindblad master equation. In the opposite limit $\alpha\rightarrow 1$, the system is confined to a given $S^z$ sector, with dynamics fully governed by the non-Hermitian Hamiltonian. 

Consider a system initialized in an unentangled state in the sector $S^z=1/2$ with parameters such that $\gamma_0 = 0$. We solve the dynamics analytically and find that the fidelity of the final $W$ state is $\mathcal{F} = 1/(3-2\alpha)$. 
Thus, as also shown in Fig.~\ref{fig:3}(b), the final tripartite entanglement increases with the level of post-selection. This behavior emerges because post-selection suppresses decay rates, and steers the system dynamics toward that induced by $H_{\mathrm{non}}$. Within the non-Hermitian dynamics, the imaginary components of the eigenvalues dictate how fast an eigenstate decays. While the imaginary part of the eigenvalues of the chiral $W$ states is $-{3a}/{2}$, the eigenvalue for the achiral $W$ state has a vanishing imaginary component and remains immune to decay. 
This mechanism increases the relative population of the dark state and causes a larger tripartite entanglement.

{This increase comes at the expense of a lower success rate for the generation of states with tripartite entanglement.
For the ideal case $|A|/a=0.5$, the success rate initially decays exponentially from 1. Subsequently, it saturates at the dark state fidelity $\mathcal{F}$ of order 10--30\%, which is not particularly small~\cite{Pfaff2014_Unconditional}. Departing from the ideal case, the fidelity $\mathcal{F}$ should be somewhat smaller due to decay from the dark state. Yet, as long as the decay rate of the dark state is much smaller than the other decay rates, there should be a time window where the fidelity (and success rate) is of similar order. This shows that the success rate is not severely suppressed, and post-selection remains a promising route to enhance tripartite entanglement. 
}

Another strategy for high fidelity generation of a $W$ state is to initialize the system in the ground state $\ket{\uparrow\uparrow\uparrow}$ and selectively drive transitions toward a specific $W$ state, while the spatially correlated noise suppresses dissipative decay. In the absence of environmentally induced coherent coupling, energy levels within a given $S^z$-sector are degenerate, while the qubit splitting $\Delta$ splits states with different $S^z$. 
As shown in Fig.~\ref{fig:3}(c), a nonzero coherent coupling $\mathcal{J}$ 
introduces a relative shift $3 \mathcal{J}$ between the chiral and achiral $W$ states within the sectors $S^z = \pm 1/2$, while maintaining a fixed energy difference between states in different total spin-$z$ sectors. 
This coherent coupling enables selective driving of an effective two-level system, consisting of $\ket{\uparrow\uparrow\uparrow}$ and a target $W$ state. Specifically, to drive transitions to the achiral $W$ state, we consider the drive Hamiltonian
\begin{equation} H_{\mathrm{drive}} = |C|(J_{\mathrm{drive}} + J_{\mathrm{drive}}^\dag)\cos[(\Delta + 2\mathcal{J})t], \end{equation} 
with $J_{\mathrm{drive}} = \sum_{i=1}^3\sigma_i^+$ and driving amplitude $|C|$ comparable to the decay rate $a$. Considering $\mathcal{J}/\Delta = 0.1$ and applying a $\pi/2$ pulse, the 
time-dependence of the resulting tripartite entanglement is shown in  Fig.~\ref{fig:3}(d), demonstrating the generation of a highly entangled state with over 99\% fidelity in the case of $|A|/a = 0.5$. For weaker correlated noise, the system exhibits reduced entanglement due to leakage back into the ground state $\ket{\uparrow\uparrow\uparrow}$. 
However, the driving protocol still enhances the entanglement significantly compared to passive evolution under the same conditions.

{We remark that the two approaches to enhance the tripartite entanglement are complementary. The coherent drive provides an active and deterministic scheme, but requires precise frequency control. Post-selection 
does not require additional control fields and is simpler to implement, but its probabilistic nature reduces the overall success rate.}

\section{Discussion}The essential requirement of our proposal is controlled balance between local and non-local noise~\cite{krauter2011entanglement}, which determines the relative size of the effective jump operators. Notably, sizable correlated noise has  been reported in gate-defined quantum dots. The strength varies with inter-dot separation, and may reach values comparable to the local noise~\cite{rojas2025inferring,Yoneda2023Noise-correlationSilicon, Rojas-Arias2023SpatialQubits}. Moreover, triangular triple-dot geometries have also been studied experimentally~\cite{Seo2013,Barthelemy2013,Flentje2017,russ2017three, Noiri2017,Acuna2024,hsieh2012physics}, and in these systems, the phase of the correlated noise is relevant.
Although experimental aspects such as tuning the phase of the noise and optimizing the triangular layout   merit further experimental investigation,  the rapid progress in quantum dot platforms indicates that the ingredients required by our proposal are already emerging in current devices. This makes the implementation of our scheme a realistic goal. {Beyond gate-defined quantum dots, our scheme is also applicable to platforms in which complex cross-correlations can be engineered, such as superconducting qubits coupled to a shared transmission-line (waveguide) environment~\cite{VonLupke2020Two-QubitQubits}.}


\begin{acknowledgements}
J.Z. thanks Shu Zhang and Yaroslav Tserkovnyak for insightful discussions.
This work was supported by the Georg H. Endress Foundation and by the Swiss National Science Foundation, NCCR SPIN (grant number 51NF40-180604). S.D. was grateful for the support by the NCCR SPIN Mobility grant. 
D.L. acknowledges the Deanship of Research and the Quantum Center at KFUPM for the support received under Grant no. CUP25102 and no. INQC2500, respectively.
\end{acknowledgements}

\appendix
\section*{APPENDIX}
\renewcommand{\topfraction}{0.95}
\renewcommand{\bottomfraction}{0.7}
\renewcommand{\textfraction}{0.03}
\renewcommand{\floatpagefraction}{0.8}
\renewcommand{\dbltopfraction}{0.95}
\renewcommand{\dblfloatpagefraction}{0.8}
\setcounter{topnumber}{4}
\setcounter{bottomnumber}{2}
\setcounter{totalnumber}{6}
\setcounter{dbltopnumber}{4}

In the following appendices, we present in Appendix~\ref{sec:master_eq} the microscopic derivation of the master equation for the three-qubit system, in Appendix~\ref{sec:analytical_sol} its analytical solutions at zero temperature, in Appendix~\ref{sec:tripartite_negativity} the tripartite negativity as a measure for the genuine tripartite entanglement, in Appendix~\ref{sec:non-homogeneous} the robustness of the entanglement dynamics against imperfections in the environment and control, and in Appendix~\ref{sec:finite_temp} the entanglement dynamics at finite temperature.

\section{Microscopic Derivation of the Master Equation for a Three-Qubit System} \label{sec:master_eq}
In this appendix we outline the derivation of the master equation for a three-qubit system interacting with a generic noisy medium. The system Hamiltonian is given by $H_S = -(\Delta/2)\sum_{\alpha=1}^{3}\,\sigma_{\alpha}^{z}$, with a global magnetic field $\vec{B} = \Delta \hat{z}$. The qubit-environment interaction is described by $H_{SE} = \lambda\sum_{\alpha=1}^{3}\left(\sigma^{+}_{\alpha}E^{-}_{\alpha} + \sigma^{-}_{\alpha}E^{+}_{\alpha}\right)$, where $\sigma^{\pm} = (\sigma^x \pm i\sigma^y)/2$, and $E^{\pm}$ are operators acting on the environment.
In the interaction picture, the Hamiltonian becomes  
\begin{equation}
    H_{SE}(-\tau) = \lambda\sum_{\alpha=1}^{3}\left[\sigma^{+}_{\alpha}E^{-}_{\alpha}(-\tau)e^{i\Delta \tau} + \sigma^{-}_{\alpha}E^{+}_{\alpha}(-\tau)e^{-i\Delta \tau}\right].
\end{equation}
\begin{widetext}
By using the following Redfield master equation~\cite{Breuer2007TheSystems,Lidar2019LectureSystems,Manzano2020AEquation}
\begin{equation}
    \frac{d\rho(t)}{dt} = -\frac{i}{\hbar}\, [H_S, \rho(t)] - \frac{1}{\hbar^2}\int_{0}^{\infty} d\tau \; \mathrm{Tr}_E \Bigl\{ \bigl[ H_{SE}, \bigl[ H_{SE}(-\tau),\, \rho(t) \otimes \rho_E^{\text{eq}} \bigr] \bigr] \Bigr\}
    \label{eq:master_equation}
\end{equation}
 and subsequently making the secular approximation, we find that 
\begin{align}
    \frac{d\rho(t)}{dt} = -\frac{i}{\hbar}\, [H_S, \rho(t)] - \frac{\lambda^2 }{\hbar^2}\sum_{\alpha,\beta=1}^{3} \int_0^\infty d\tau\, &\Big\{ \langle E^{-}_{\alpha} E^{+}_{\beta}(-\tau)\rangle\, e^{-i\Delta \tau} \Big(\sigma^{+}_{\alpha}\sigma^{-}_{\beta}\rho(t) -\sigma^{-}_{\beta}\rho(t)\sigma^{+}_{\alpha}\Big) \nonumber\\
    & +\langle E^{+}_{\alpha} E^{-}_{\beta}(-\tau)\rangle\, e^{i\Delta \tau} \Big(\sigma^{-}_{\alpha}\sigma^{+}_{\beta}\rho(t) -\sigma^{+}_{\beta}\rho(t)\sigma^{-}_{\alpha}\Big) + \text{H.c.} \Big\} .
\end{align}
We regroup the terms and write the master equation in the following Lindblad form: 
\begin{align}
    \frac{d\rho(t)}{dt} = -\frac{i}{\hbar}\, [H_S + H_{LS}, \rho(t)] +\sum_{\alpha,\beta=1}^{3} \Big\{&\tilde{\gamma}_{\alpha\beta} \Big[\sigma^{-}_{\beta}\rho(t)\sigma^{+}_{\alpha}-\frac{1}{2}\{\sigma^{+}_{\alpha}\sigma^{-}_{\beta},\rho(t)\}\Big]  
     +\gamma_{\alpha\beta} \Big[\sigma^{+}_{\beta}\rho(t)\sigma^{-}_{\alpha} - \frac{1}{2}\{\sigma^{-}_{\alpha}\sigma^{+}_{\beta},\rho(t)\}\Big]\Big\}.
\end{align}
where we introduced the dissipation rates 
\begin{align}
    \gamma_{\alpha\beta} &= \frac{\lambda^2}{\hbar^2}\int_{0}^{\infty} d\tau\, \langle E^{+}_{\alpha} E^{-}_{\beta}(-\tau)\rangle\, e^{i\Delta \tau} + \frac{\lambda^2}{\hbar^2}\int_{-\infty}^{0} d\tau\, \langle E^{+}_{\alpha} E^{-}_{\beta}(-\tau)\rangle\, e^{i\Delta \tau} = \frac{\lambda^2}{\hbar^2}\int_{-\infty}^{\infty} d\tau\, \langle E^{+}_{\alpha} E^{-}_{\beta}(-\tau)\rangle\, e^{i\Delta \tau},\\
    \tilde{\gamma}_{\alpha\beta} &= \frac{\lambda^2}{\hbar^2}\int_{0}^{\infty} d\tau\, \langle E^{-}_{\alpha} E^{+}_{\beta}(-\tau)\rangle\, e^{-i\Delta \tau} + \frac{\lambda^2}{\hbar^2}\int_{-\infty}^{0} d\tau\, \langle E^{-}_{\alpha} E^{+}_{\beta}(-\tau)\rangle\, e^{-i\Delta \tau} = \frac{\lambda^2}{\hbar^2}\int_{-\infty}^{\infty} d\tau\, \langle E^{-}_{\alpha} E^{+}_{\beta}(-\tau)\rangle\, e^{-i\Delta \tau}.
    \label{eq_gammaDef}
\end{align}
and the effective coherent interaction 
\begin{equation}
    H_{LS} = \sum_{\alpha,\beta=1}^3 (\mathcal{J}_{\alpha\beta} \sigma_\alpha^-\sigma_\beta^+ + \bar{\mathcal{J}}_{\alpha\beta} \sigma_\alpha^+\sigma_\beta^- ),\;\; 
    \label{eq_H_LS}
\end{equation}   
with in-plane exchange coupling strengths 
\begin{align}
    \mathcal{\bar{J}}_{\alpha\beta} &= - i \frac{\lambda^2}{2\hbar^2}\int_0^\infty d\tau \bigg( \langle E^{-}_{\alpha} E^{+}_{\beta}(-\tau)\rangle e^{-i\Delta \textstyle \tau} - \langle E^{-}_{\alpha}(-\tau) E^{+}_{\alpha}\rangle e^{i\Delta \textstyle \tau}  \bigg), \\
    \mathcal{J}_{\alpha\beta} &= -i \frac{\lambda^2}{2\hbar^2 } \int_0^\infty d\tau \bigg( \langle E^{+}_{\alpha} E^{-}_{\beta}(-\tau)\rangle e^{i\Delta \textstyle \tau} - \langle E^{+}_{\alpha}(-\tau) E^{-}_{\alpha}\rangle e^{-i\Delta \textstyle \tau}  \bigg).
\end{align}
\end{widetext}
In the following, we discuss the Lindbladian and the induced coherent interaction in further detail. \\

\textbf{\textit{Induced Coherent Interaction}}|
The induced coherent interaction $H_{LS}$ can be decomposed into a local part $H_{LS}^\mathrm{loc}$ arising from terms with $\alpha = \beta$, and a nonlocal part $H_\mathrm{eff}$ due to terms with $\alpha \neq \beta$. The local contribution results in a frequency shift of the bare qubit splitting, as seen from:
\begin{equation}
\al{    H_{LS}^\mathrm{loc} &=  \sum_{\alpha} \left( \mathcal{J}_{\alpha\alpha}\sigma^+_\alpha\sigma^-_\alpha + \text{H.c.} \right)\\
    &= \frac{ \lambda^2}{2}\sum_{\alpha}\,\text{Im} \int_0^\infty \!\! d\tau \, e^{-i\Delta \tau} \langle \{E^{-}_{\alpha}(\tau), E^{+}_{\alpha}\} \rangle {\sigma^z_\alpha},}
\end{equation}
where we have omitted a constant term. This Lamb shift can be absorbed into a redefinition of the qubit splitting, and we neglect it in the subsequent discussion. We therefore only consider $H_\mathrm{eff}$, which describes environmentally induced coherent interactions between qubits. It is straightforward to show that it takes the form: 
\begin{equation}
\al{    H_{\text{eff}} = 
     \sum_{\substack{\alpha = 1,\\\beta=\alpha+1}}^3 \big[&(\mathcal{J}^{\alpha\beta}_\perp + iD^{\alpha\beta}) \sigma^+_\alpha\sigma^-_\beta\\ &+ (\mathcal{J}^{\alpha\beta}_\perp - iD^{\alpha\beta}) \sigma^-_\alpha\sigma^+_\beta\big],}
\end{equation}
where the symmetric and antisymmetric coupling coefficients are given by
\begin{align}
    \mathcal{J}^{\alpha\beta}_\perp &= \,\,\,\,\,\,\, \frac{\lambda^2}{4} \big[ 
    G^A_{E^+_\beta E^-_\alpha}(\Delta) + G^R_{E^+_\beta E^-_\alpha}(\Delta)\nonumber\\
    &\qquad\quad+ G^A_{E^+_\alpha E^-_\beta}(\Delta) + G^R_{E^+_\alpha E^-_\beta}(\Delta) \big], \\
    D^{\alpha\beta} &= -i \frac{\lambda^2}{4} \big[
    G^A_{E^+_\beta E^-_\alpha}(\Delta) + G^R_{E^+_\beta E^-_\alpha}(\Delta)\nonumber\\
    &\qquad\quad- G^A_{E^+_\alpha E^-_\beta}(\Delta) - G^R_{E^+_\alpha E^-_\beta}(\Delta) \big].
\end{align}
Here, we use the standard definitions of the retarded and advanced Green’s functions:
\begin{equation}
\al{    G^R_{AB}(\Delta) &= -i\, \int_0^{\infty}d\tau \, e^{i\Delta\tau} \, \langle [A(\tau), B(0)]\rangle \;\;\text{and}\\ G^A_{AB}(\Delta) &= i \int_{-\infty}^0 dt\, e^{i\Delta t}\, \langle [A(t),B(0)]\rangle.}
\end{equation}
The Hamiltonian can also be rewritten as 
\begin{equation}
\al{    H_{\text{eff}} = \sum_{\substack{\alpha = 1 \\ \beta = \alpha+1}}^3 \Big[&
    \frac{\mathcal{J}^{\alpha\beta}_\perp}{2}(\sigma_\alpha^x \sigma_\beta^x + \sigma_\alpha^y \sigma_\beta^y)\\
    &+ \frac{D^{\alpha\beta}}{2} \hat{z} \cdot (\vec{\sigma}_\alpha \times \vec{\sigma}_\beta)
    \Big],}
\end{equation}
and clearly corresponds to an in-plane exchange interaction and out-of-plane DMI. 

We remark that the induced coupling arises purely from quantum noise correlations in the environment. This is evident from the fact that the coupling coefficients are expressed in terms of anticommutators of the environmental operators, which directly encode the quantum correlations of the noise.\\

\textbf{\textit{Lindbladian}}|The Lindbladian can be written in the form
\begin{align}
    \mathcal{L}[\rho] = &\sum_{\alpha\beta}\gamma_{\alpha\beta}\big(\mathcal{O}_\beta\rho\mathcal{O}^\dag_\alpha - \frac{1}{2}\{\mathcal{O}_\alpha^\dag\mathcal{O}_\beta,\rho\}\big)\nonumber\\
    &+ \sum_{\alpha\beta}\tilde{\gamma}_{\alpha\beta}\big(\tilde{\mathcal{O}}_\beta\rho\tilde{\mathcal{O}}^\dag_\alpha - \frac{1}{2}\{\tilde{\mathcal{O}}_\alpha^\dag\tilde{\mathcal{O}}_\beta,\rho\}\big),
\end{align}
with the operator basis $\mathcal{O} = \{\sigma_1^+,\sigma_2^+,\sigma_3^+\}$ and $\tilde{\mathcal{O}} = \{\sigma_1^-,\sigma_2^-,\sigma_3^-\}$. The environmental correlators in Eq.~\eqref{eq_gammaDef} can be expressed in terms of power spectral densities
\begin{equation}
    \label{eq:PSD}
    S_{ij}(\Delta) = \int_{-\infty}^{\infty} dt\, e^{i\Delta t}\langle E^+_i(t)E^-_j\rangle,
\end{equation}
which connect the dissipative coefficients directly to measurable spectral properties of the environment.
Then, one can immediately observe that $\gamma_{\alpha\beta} = \lambda^2S_{\alpha\beta}(\Delta)/\hbar^2$. By invoking the fluctuation-dissipation theorem, one may show that
  $\tilde{\gamma}_{\beta\alpha}  = \exp(-\beta\hbar\Delta)\gamma_{\alpha\beta}$. Assuming a homogeneous medium, such that every qubit experiences the same noise, we have $\gamma_{\alpha\alpha} = a$ with real valued local dissipation $a$, and non-local dissipation $\gamma_{12}=\gamma_{23}=\gamma_{31} = A$, which is generally complex valued. For the reverse processes, we introduce $\gamma_{\alpha\alpha} = \tilde{a}$ and $\gamma_{13}=\gamma_{32}=\gamma_{21} = \tilde{A}$, such that $\tilde{A} = \exp(-\beta\hbar\Delta) A$ and $\tilde{a} = \exp(-\beta\hbar\Delta) a$. As a result we have the two transition rate matrices
\begin{equation}
\al{    \gamma &= 
    \begin{pmatrix}
        a & A & A^* \\
        A^* & a & A \\
        A & A^* & a 
    \end{pmatrix},\\
    \tilde{\gamma} &= 
    \begin{pmatrix}
        \tilde{a} & \tilde{A}^* & \tilde{A} \\
        \tilde{A} & \tilde{a} & \tilde{A}^* \\
        \tilde{A}^* & \tilde{A} & \tilde{a} 
    \end{pmatrix}.}
\end{equation}
where $\gamma$ is the probability rate related to decay processes in the system and $\tilde{\gamma}$ is the rate related to excitations in the system. 
To diagonalize the Lindbladian, we introduce jump operators $J_k$ and $\tilde{J}_k$ through the transformation $\mathcal{O}_k = \sum_l U_{kl} J_{l}$ 
and $\tilde{\mathcal{O}}_k = \sum_l \tilde{U}_{kl} \tilde{J}_l$, 
where we choose the matrices $U_{kl}$ and $\tilde{U}_{kl}$ such that $U^\dagger \gamma U$ and $\tilde{U}^\dagger \tilde{\gamma} \tilde{U}$ are diagonal, while $U^\dagger U \propto 1$ and $\tilde{U}^\dagger \tilde{U} \propto 1$ (i.e. unitary up to overall normalization). Choosing the normalization of $U$ and $\tilde{U}$ appropriately,
the Lindbladian then takes the form $\mathcal{L}[\rho] = \sum_{k} ( \mathcal{D}_{J_k}[\rho] + \mathcal{D}_{\tilde{J}_k}[\rho])$ with $\mathcal{D}_{J_k} = J_k \rho J_k^\dag - \frac{1}{2}\{J_k^\dag J_k,\rho\}$. The operators $J_k$ are the jump operators that take the system from one eigenstate of the full unitary dynamics ($H_S+H_{\mathrm{eff}}$) to another, and are given by 

\begin{equation}
    J_k = \sqrt{\frac{\gamma_k(a,A)}{3}}(\eta^k\sigma_1^+ + \eta^{-k}\sigma_2^+ + \sigma_3^+),
    \label{eq:J_k_app}
\end{equation}
where  $k\in \{-1,0,1\}$, $\eta = \exp(2\pi i/3)$, and the decay rates are the eigenvalues of the matrix $\gamma$. After introducing the notation that $A=|A|e^{i\phi}$, these decay rates are given by
\begin{equation}
    \gamma_k(a,A) =  {a + 2|A|\cos{(\phi +2\pi k / 3)}}.
    \label{eq:gamma_k_app}
\end{equation}
Since the matrix $\tilde{\gamma}$ is proportional to the matrix $\gamma$, it is diagonalized by the same transformation. The jump operators therefore assume the same form
\begin{equation}
    \tilde{J}_k = \sqrt{\frac{\gamma_k(\tilde{a},\Tilde{A})}{3}}(\eta^{-k}\sigma_1^- + \eta^{k}\sigma_2^- + \sigma_3^-),
\end{equation}
where the jump amplitude has been modified according to $a \rightarrow \tilde{a}$ and $A \rightarrow \tilde{A}$.

A feature that is distinct from the 2-qubit case \cite{Zou2022Bell-stateCoupling}, is the appearance of the phase $\phi$ in the decay rates. This arises because the phase cannot be absorbed into operators $\sigma_i^\pm$ (i.e. gauged away) due to the triangular geometry.

\reapp{  We finally remark that our analysis has focused on transverse qubit–environment coupling. When the dominant coupling is longitudinal, that is, when the total Hamiltonian takes the form
\( H= - \frac{\Delta}{2} \sum_\alpha \sigma_\alpha^z  + \lambda \sum_\alpha \sigma_\alpha^z E_\alpha +H_E,   \) 
we can effectively rotate the qubit quantization axis such that the longitudinal coupling becomes effectively transverse through a resonance drive~\cite{Zou2024SpatiallyQubits}. To see this, assuming we have a drive $\sum_\alpha\Omega\cos(\Delta t/\hbar)\sigma_\alpha^x$, the Hamiltonian in the rotating frame then is given by}
\(  \al{ &\mathcal{H}=i(\hbar\partial_t R)R^\dagger +RHR^\dagger, \\ &\text{with}\;\;\; R(t)=\exp(-\frac{i\Delta t}{2\hbar} \sum_{\alpha} \sigma_\alpha^z  ). } \)
\reapp{We note that the unitary operator $R(t)$ commutes with the coupling $\lambda \sum_\alpha \sigma_\alpha^z E_\alpha$  and $H_{E}$, leaving them invariant in the rotating frame. After
applying the rotating-wave approximation by neglecting counter-rotating
terms that oscillate fast at frequency $2\Delta/\hbar$, the total Hamiltonian becomes $\mathcal{H}=\sum_{\alpha}\Omega \sigma_\alpha^x/2 + \lambda \sum_\alpha \sigma_\alpha^z E_\alpha +H_E$. Thus, the qubit-environment coupling effectively becomes transverse. One can further rotate the spin axis in spin space and write the Hamiltonian as $\mathcal{H}= - (\Omega/2)\sum_\alpha \sigma_\alpha^z + \lambda\sum_\alpha \sigma^x_\alpha E_\alpha +H_E$, where the environment only leads to energy relaxation. In Appendix~\ref{sec:non-homogeneous}, we also discuss in detail the entanglement dynamics in the presence of both transverse and longitudinal couplings (both energy relaxation and pure dephasing). 
}


\section{Analytical Solution of System Dynamics at Zero Temperature} \label{sec:analytical_sol}

In this appendix, we solve the master equation for the three-qubit system $[\rho_{ij}]_{8\times 8}$ analytically at zero temperature. Then, one only needs to consider three out of six jump operators, as $\tilde{a}$ and $\tilde{A}$ vanish. We solve the master equation in the eigenbasis of the full unitary dynamics. These states can be labeled by their total spin, $S^z$, and the chirality eigenvalue, $\chi$. This basis consists of the states $\ket{S^z=3/2,\chi=0}=\ket{\uparrow\uparrow\uparrow}$,
$\ket{S^z=-3/2,\chi=0}=\ket{\downarrow\downarrow\downarrow}$, but also of the entangled chiral states of the form 
\begin{align}
    \label{eq:eigenstate_W}
    \ket{S^z=\tfrac{1}{2},\chi} &= \frac{1}{\sqrt{3}} \left({\eta^{-\chi}\ket{\uparrow\uparrow\downarrow} + \eta^{\chi}\ket{\uparrow\downarrow\uparrow} + \ket{\downarrow\uparrow\uparrow}} \right),\\
    \ket{S^z=-\tfrac{1}{2},\chi} &= \frac{1}{\sqrt{3}} \left({\eta^{-\chi}\ket{\downarrow\downarrow\uparrow} + \eta^{\chi}\ket{\downarrow\uparrow\downarrow} + \ket{\uparrow\downarrow\downarrow}} \right).
\end{align}
Here, we have $\eta=\exp(2\pi i/3)$ and $\chi\in \{-1,0,1\}$.
To simplify notation in the density matrix evolution, we introduce the quantities 
\(  \al{  \gamma_j &= a + 2|A|\cos{\big(\phi + 2\pi j/3\big)}, \;\;\text{and}\\  f_j &= 2\mathcal{J}\cos{\big(\psi + 2\pi j/3\big)}, } \)
with $j\in \{-1,0,1\}$.
For the $\gamma$ parameters we have the property that $\sum_{i}\gamma_i = 3a$. The order of the basis in which we solve for the density matrix is $\{\ket{\uparrow\uparrow\uparrow}$, $\ket{1/2,0}$, $\ket{1/2,-1}$, $\ket{1/2,1}$, $\ket{-1/2,0}$, $\ket{-1/2,-1}$, $\ket{-1/2,1}$, $\ket{\downarrow\downarrow\downarrow}\}$ where we used the notation $\ket{S^z,\chi}$ for all $W$ states. The density matrix takes a block diagonal structure with blocks structured according to $S^z$, such that we have two blocks of a single element corresponding to $S^z = \pm 3/2$ and two blocks of 3$\times$3 matrices corresponding to $S^z=\pm1/2$. We denote the blocks by a superscript containing its corresponding sector as $\rho^{S^z}_{ij}$. One can also introduce matrices containing the amplitudes of the jumps between one sector and another. These are the matrices $\Upsilon$ where the superscript denotes the sectors between the jumps occur. Its basis is given in terms of chirality in the order $\chi \in \{0,-1,1\}$, as they describe $ \Upsilon_{ij}= |\langle S^z = S+1, \chi=i | \sum_k J_k | S^z=S, \chi = j\rangle |^2$. For the jump between sector $S^z=-3/2$ and $S^z=-1/2$ we obtain that 
\begin{equation}
    \Upsilon^{3\rightarrow2} = \begin{pmatrix}
        \gamma_0  \\
         \gamma_{-1} \\
         \gamma_{1}
    \end{pmatrix}.
\end{equation}    
For the jumps between sector $S^z=-1/2$ and $S^z=1/2$, the matrix is 
\begin{equation}
    \Upsilon^{2\rightarrow1} = \frac{1}{3}
     \begin{pmatrix}
        4\gamma_0 & \gamma_{1} & \gamma_{-1} \\
        \gamma_{-1} & \gamma_0 & 4\gamma_{1} \\
        \gamma_{1} & 4\gamma_{-1} & \gamma_0\\
     \end{pmatrix}.
\end{equation}
Next we also introduce some matrices related to exponential decay rates which will be useful in the following
\( \setlength{\arraycolsep}{2.5pt} \al{   \Gamma^{3\rightarrow2} &=
    \begin{pmatrix}
        \gamma_0 - 2a\\
        \gamma_{-1} - 2a  \\
      \gamma_{1} - 2a
    \end{pmatrix}, \\ \Gamma^{2\rightarrow1} &= 
     \begin{pmatrix}
        -a & \gamma_0-\gamma_{-1} - a & \gamma_0 - \gamma_{1} - a \\
        \gamma_{1} - \gamma_0 - a & \gamma_{1} - \gamma_{-1} - a &-a \\
        \gamma_{-1} - \gamma_0 -a & -a & \gamma_{-1} - \gamma_{1} - a
     \end{pmatrix}.} \)

We will denote the different density matrix elements with a superscript denoting the total $S^z$ sector in the system, and as subscript the matrix element within the sector. For notational convenience we set here $\hbar = 1$.

\textbf{\textit{Initial condition $S^z = 1/2$.}} We will start with solution to the density matrix elements when we consider our initial density matrix to be a pure spin-state in the sector $S^z=1/2$. With the parameters introduced above, the density matrix elements become 
\begin{equation}
\al{    \rho_{ij}^{S^z=1/2}(t) = \rho_{ij}^{S^z=1/2}(0)&\exp[-(\gamma_{i}+\gamma_{j})t/2]\\ &\times\exp[i(f_{i}-f_j)t],}
\end{equation}
where indices indicate the chirality in the order $\{0,1,-1\}$, and $ \rho^{S^z=3/2}(t) = 1-\operatorname{Tr} \rho^{S^z=1/2}(t),$ following directly from the constraint $\operatorname{Tr} \rho=1$.

\textbf{\textit{Initial condition $S^z = -1/2$.}} Next we can go one step up and start in the $S^z = -1/2$ sector. The solutions for the density matrix elements corresponding to each of the sectors labeled by $S^z$ are presented below. First, the matrix elements in the sector $S^z = -1/2$ are given by 
\begin{align}
    \rho^{S^z=-1/2}_{11}(t) &= \rho^{S^z=-1/2}_{11}(0) \exp[-(a+\gamma_0) t],\\
    \rho^{S^z=-1/2}_{22}(t) &= \rho^{S^z=-1/2}_{22}(0) \exp[-(a+\gamma_{-1})t],\\
    \rho^{S^z=-1/2}_{33}(t) &= \rho^{S^z=-1/2}_{33}(0) \exp[-(a+\gamma_{1})t], \\
    \rho^{S^z=-1/2}_{12}(t) &= \rho^{S^z=-1/2}_{12}(0)\nonumber\\ &\hspace{-6pt}\times \exp[-(2a + \gamma_0 + \gamma_{-1})t/2]\exp[i(f_0 - f_{-1})t],\\
    \rho^{S^z=-1/2}_{13}(t) &= \rho^{S^z=-1/2}_{13}(0)\nonumber\\ &\hspace{-6pt}\times \exp[-(2a + \gamma_0 + \gamma_{1})t/2]\exp[i(f_0 - f_{1})t],\\
    \rho^{S^z=-1/2}_{23}(t) &= \rho^{S^z=-1/2}_{23}(0)\nonumber\\ &\hspace{-6pt}\times \exp[-(2a + \gamma_{-1} + \gamma_{1})t/2]\exp[i(f_{-1} - f_{1})t].
\end{align}
For the sector $S^z=1/2$, we have occupations
\begin{equation}
  \al{  \rho_{ii}^{S^z=1/2}(t) &= \sum_{j=1}^{3}\frac{\Upsilon^{2\rightarrow1}_{ij}}{\Gamma^{2\rightarrow1}_{ij}} \big[1 - \exp(-\Gamma^{2\rightarrow1}_{ij}t)\big] \rho^{S^z=-1/2}_{jj}(t),\\
      }
\end{equation}\\
while the off-diagonal terms are given by: 
\begin{widetext}
\begin{align}
  \rho_{12}^{S^z=1/2}(t) = \frac{1}{3}\Bigg\{ &
    -\rho_{12}^{S^z=-1/2}(t)\,
    \frac{2\gamma_{0}\bigl[1 - e^{(a - \gamma_{1}/2 + \gamma_{-1}/2)t - i(f_{-1}-f_{1})t}\bigr]}
         {-a + \gamma_{1}/2 - \gamma_{-1}/2 + i(f_{-1}-f_{1})}
    + \rho_{31}^{S^z=-1/2}(t)\,
    \frac{\gamma_{-1}\bigl[1 - e^{at - 2i(f_{-1}-f_0)t}\bigr]}
         {-a + 2i(f_{-1}-f_0)}
    \notag\\
  &\quad -\,\rho_{23}^{S^z=-1/2}(t)\,
    \frac{2\gamma_{1}\bigl[1 - e^{(a - \gamma_{0}/2 + \gamma_{-1}/2)t - i(f_{1}-f_0)t}\bigr]}
         {-a + \gamma_{0}/2 - \gamma_{-1}/2 + i(f_{1}-f_0)}
  \Bigg\},
\end{align}

\begin{align}
  \rho_{13}^{S^z=1/2}(t) = \frac{1}{3}\Bigg\{ &
    -\rho_{13}^{S^z=-1/2}(t)\,
    \frac{2\gamma_{0}\bigl[1 - e^{(a - \gamma_{-1}/2 + \gamma_{1}/2)t - i(f_{1}-f_{-1})t}\bigr]}
         {-a + \gamma_{-1}/2 - \gamma_{1}/2 + i(f_{1}-f_{-1})}
    - \rho_{32}^{S^z=-1/2}(t)\,
    \frac{2\gamma_{-1}\bigl[1 - e^{(a - \gamma_{0}/2 + \gamma_{1}/2)t - i(f_{-1}-f_0)t}\bigr]}
         {-a + \gamma_{0}/2 - \gamma_{1}/2 + i(f_{-1}-f_0)}
    \notag\\
  &\quad +\,\rho_{21}^{S^z=-1/2}(t)\,
    \frac{\gamma_{1}\bigl[1 - e^{at - 2i(f_{1}-f_0)t}\bigr]}
         {-a + 2i(f_{1}-f_0)}
  \Bigg\}.
\end{align}

\begin{align}
    \rho_{23}^{S^z=1/2}(t) = \frac{1}{3}\Bigg\{&  \rho_{23}^{S^z=-1/2}(t) \frac{\gamma_0 \big[1 - e^{at + 2i(f_{1} - f_{-1}) t} \big]}{-a + 2i(f_{1} - f_{-1})}   
    - \rho_{12}^{S^z=-1/2}(t) \frac{2\gamma_{-1} \big[ 1 - e^{(a-\gamma_{1}/2 + \gamma_0/2)t+i(f_0 - f_{-1}) t} \big] }{(-a+\gamma_{1}/2 - \gamma_0/2) +i(f_0 - f_{-1})}  \notag \\ 
    &- \rho_{31}^{S^z=-1/2}(t) \frac{2\gamma_{1} \big[1 - e^{(a-\gamma_{-1}/2+\gamma_0/2)t  + i(f_{1} - f_0) t}\big] }{(-a+\gamma_{-1}/2-\gamma_0/2) +i(f_{1} - f_0)} \Bigg\}.
\end{align}
\end{widetext}
Finally, the element in the block corresponding to $S^z = 3/2$ is given by 
\begin{align}
\rho^{S^z=3/2}(t) &= 1-\text{Tr}[\rho^{S^z=1/2}(t)]-\text{Tr}[\rho^{S^z=-1/2}(t)].
\end{align}
Altogether, this completely specifies the time evolution of the density matrix. 

\textbf{\textit{Initial condition $S^z = -3/2$.}} Initializing from this state means that one cannot reach any off-diagonal elements in the density matrix in the Hamiltonian eigenbasis as jump operators only take you from a diagonal term to another diagonal one.
For the initial state $\ket{\downarrow\downarrow\downarrow}$, the density matrix element evolves in time according to 
\begin{equation}
\al{    \rho^{S^z=-3/2}(t) &= \exp [-(\gamma_0 + \gamma_{-1} + \gamma_{1})t] \rho^{S^z=-3/2}(0)\\ &= e^{-3at} \rho^{S^z=-3/2}(0).}
\end{equation}
Within the sector $S^z=-1/2$, the non-zero matrix elements are given by
\begin{equation}
    \rho_{ii}^{S^z=-1/2}(t) =  \frac{\Upsilon^{3\rightarrow2}_{i}}{\Gamma^{3\rightarrow 2}_{i}} \big[ 1 - \exp(-\Gamma^{3\rightarrow 2}_{i} t) \big] \rho^{S^z=-3/2}(t).
\end{equation}
Now we get for the diagonal elements in the $S^z=1/2$ sector using the matrices introduced above that

\begin{widetext}
\begin{align}
    \rho_{ii}^{S^z=1/2}(t) &= \sum_{j=1}^{3}\frac{\Upsilon^{3\rightarrow2}_{j}\Upsilon^{2\rightarrow1}_{ij}}{\Gamma^{3\rightarrow2}_{j}\Gamma^{2\rightarrow1}_{ij}}\bigg[1 - \exp(- \Gamma_{j}^{3\rightarrow2}t)
    - \frac{\Gamma_{j}^{3\rightarrow2}}{\Gamma_{j}^{3\rightarrow2} + \Gamma_{ij}^{2\rightarrow1}}\big\{ 1 - \exp[-(\Gamma_{j}^{3\rightarrow2}+\Gamma^{2\rightarrow 1}_{ij})t]\big\}  \bigg] \rho^{S^z=-3/2}(t),
\end{align}
\end{widetext}
and finally
\begin{equation}
\al{    \rho^{S^z=3/2}(t) = 1 &- \text{Tr}[\rho^{S^z=1/2}(t)]\\ &- \text{Tr}[\rho^{S^z=-1/2}(t)]- \rho^{S^z=-3/2}(t).}
\end{equation}
With these explicit solutions for the time evolution of the three-qubit density matrix, we are ready to extract the dynamics of the tripartite entanglement in the system for different initial conditions.

\section{Tripartite Negativity as a Tripartite Entanglement Measure} \label{sec:tripartite_negativity}
To study the entanglement dynamics of the three spin qubit system described by the Lindblad equation one needs an entanglement measure which is applicable to both pure and mixed states. The negativity defined by Vidal et al. is a measure that can do this \cite{Vidal2002ComputableEntanglement}.
It is defined via the partial transpose $T_j$ of the density matrix $\rho$ with respect to qubit $j$.
For a density matrix which in the Pauli basis takes the form
\begin{align}
\rho = \sum_{\{\sigma_j\}, \{\rho_j\}} \rho_{\sigma_1\sigma_2\sigma_3; \rho_1\rho_2\rho_3} |\sigma_1 \sigma_2 \sigma_3 \rangle \langle \rho_1 \rho_2 \rho_3|
\end{align}
the partial transform gives e.g. $\rho^{T_3}_{\sigma_1 \sigma_2 \sigma_3;\rho_1\rho_2\rho_3} = \rho_{\sigma_1\sigma_2\rho_3; \rho_1 \rho_2 \sigma_3}$. 
The bipartite negativity is then given by
\begin{equation}
    \mathcal{N}_j = 2 \sum_i \big| \min\{0, \lambda_i^{T_j} \} \big|,
\end{equation}
where $\lambda_i^{T_j}$ are the eigenvalues of $\rho^{T_j}$.
However, when one calculates the negativity with a particular choice of $j$, one qubit is separated out explicitly from the rest of the system; thus, the resulting entanglement dynamics will be dependent on the initial condition, as well as the chosen qubit.

This measure does not in itself capture the entanglement among all three qubits, and one cannot separate the entanglement among pairs of qubits from entanglement across all three qubits. An entanglement measure that can do this is referred to as a Genuine Multipartite Entanglement (GME) measure~\cite{Ma2011MeasureBounds}. We use the geometrically averaged version of all bipartite negativities as introduced by Sabín et al.~\cite{Sabin2008ASystems},
\begin{equation}
    \mathcal{N}_{123} = \sqrt[3]{\mathcal{N}_{[1]23}\mathcal{N}_{1[2]3}\mathcal{N}_{12[3]}},
    \label{eq_tripartiteNegativity}
\end{equation}
which naturally makes this distinction since it will be zero if any of the bipartite negativities is zero. Second, to qualify as a GME measure, it must also rank the GHZ state ($\ket{\mathrm{GHZ}} = \tfrac{1}{\sqrt{2}} \{\ket{\uparrow\uparrow\uparrow} - \ket{\downarrow\downarrow\downarrow}\}$) as more entangled than the $W$ state ($\ket{\mathrm{W}} = \tfrac{1}{\sqrt{3}} \{\ket{\uparrow\downarrow\downarrow} + \ket{\downarrow\uparrow\downarrow} + \ket{\downarrow\downarrow\uparrow}\}$)~\cite{Xie2021TriangleEntanglement, Ma2024MultipartiteReview}. The measure in Eq.~\eqref{eq_tripartiteNegativity} also fulfills this requirement, as $\mathcal{N}_{123}(\mathrm{GHZ}) = 1$ and $\mathcal{N}_{123}(\mathrm{W}) = 2\sqrt{2}/3 \approx 0.94$. Since the above measure is fairly easy to compute, can be applied to both mixed and pure states, and captures GME, it is an effective metric to study entanglement dynamics in multipartite systems.

\textit{\textbf{Calculation of tripartite negativity.}}
For the evolution of the density matrix considered here, in the Pauli basis $\{\ket{\uparrow\uparrow\uparrow},\ket{\downarrow\uparrow\uparrow},\ket{\uparrow\downarrow\uparrow}, \ket{\uparrow\uparrow\downarrow}, \ket{\uparrow\downarrow\downarrow}, \ket{\downarrow\uparrow\downarrow}, \ket{\downarrow\downarrow\uparrow}, \ket{\downarrow\downarrow\downarrow}\}$, the density matrix generally takes a block-diagonal form:
\begin{widetext}
\begin{equation}
\rho(t)= 
\begin{bmatrix}
    \rho^{S^z = 3/2}(t)  &   \mathbf{0}  &   \mathbf{0} & \mathbf{0}           \\
    \mathbf{0}      & [\rho^{S^z = 1/2}_{ij}(t)]_{3\times 3} &   \mathbf{0} & \mathbf{0}         \\
    \mathbf{0}      &   \mathbf{0}  & [\rho^{S^z = -1/2}_{ij}(t)]_{3\times 3}  & \mathbf{0}       \\
     \mathbf{0}  &   \mathbf{0} & \mathbf{0} & \rho^{S^z = -3/2}(t)
\end{bmatrix}.
\end{equation}
\end{widetext}
The partially transposed density matrix is still block-diagonal with two 3$\times$3 matrices and two single diagonal elements. For example, when the partial transpose is taken with respect to qubit 3, the two 3$\times$3 matrices are of the form
\begin{align}
    \setlength{\arraycolsep}{2.5pt}
    M^{S^z = -1/2}_3 &= 
    \begin{bmatrix}
        \rho_{11}^{S^z = 1/2} &\rho_{12}^{S^z = 1/2} & \rho_{23}^{S^z = -1/2} \\[.3cm]
        \rho_{21}^{S^z = 1/2}& \rho_{22}^{S^z = 1/2} & \rho_{13}^{S^z = -1/2}\\[.3cm]
        \rho_{32}^{S^z = -1/2}& \rho_{31}^{S^z = -1/2}& \rho^{S^z = -3/2}
    \end{bmatrix}, \nonumber\\ M^{S^z = 1/2}_3 &= 
    \begin{bmatrix}
        \rho_{11}^{S^z = -1/2} &\rho_{12}^{S^z = -1/2} & \rho_{23}^{S^z = 1/2} \\[.3cm]
        \rho_{21}^{S^z = -1/2}& \rho_{22}^{S^z = -1/2} & \rho_{13}^{S^z = 1/2}\\[.3cm]
        \rho_{32}^{S^z = 1/2}& \rho_{31}^{S^z = 1/2}& \rho^{S^z = 3/2}
    \end{bmatrix}.\label{eq:3x3mat_qubit3}
\end{align}
The negative eigenvalues of these matrices are the ones that cause non-zero entanglement in the negativity entanglement measure. However, in our analytical solution [see Appendix \ref{sec:analytical_sol}], the density matrix is obtained in the eigenbasis of the Hamiltonian $H=H_S+H_{\mathrm{eff}}$ governing the full unitary dynamics, where $H_S$ represents the system Hamiltonian and $H_{\mathrm{eff}}$ describes the induced coherent dynamics. This eigenbasis corresponds to entangled eigenstates. Therefore, the transformation from this eigenbasis to the Pauli basis is non-local with respect to different qubits. 

When determining the time evolution of the eigenstates, it is useful to work in the eigenstate basis. Yet, this is not a convenient basis to work in when we calculate the  entanglement. Since the definition of the multipartite entanglement involves the partial transpose with respect to a given qubit, it is more convenient to work in the Pauli basis. Let $U$ denote the unitary basis transformation from the eigenbasis to the Pauli basis, so that $\rho_{\mathrm{Pauli}} = U\rho_{\mathrm{eigen}}U^\dag$. In the case where $U$ can be factorized as $U = U_1\otimes U_2 \otimes U_3 $, the partial transpose with respect to any qubit is left invariant under the basis transformation. However, in our case it is non-local---it does not factorize as $U_1\otimes U_2 \otimes U_3$---due to the entanglement of the eigenstates. For further details on the invariance properties under local unitaries and consequences of non-local transformations, we refer to Refs. \cite{Vidal2002ComputableEntanglement, Horodecki2009QuantumEntanglement}.
To calculate the multipartite entanglement, we always determine the time evolution in the eigenbasis first, and then rotate back to the Pauli basis before we determine the multipartite entanglement as described above.  

An example density matrix that arises frequently in our system is a density matrix of the form
\begin{equation}
    \rho = \mathcal{F}\ket{\mathrm{W}}\!\bra{\mathrm{W}} + (1-\mathcal{F})\ket{\uparrow\uparrow\uparrow}\!\bra{\uparrow\uparrow\uparrow},
\end{equation}
where $\ket{\mathrm{W}} = \ket{S^z=1/2,\chi=0}$ as given in Eq.~\eqref{eq:eigenstate_W}. It is straightforward to write out this equation in the Pauli basis and, with the procedure outlined above, one can calculate the tripartite negativity in terms of the $W$-state fidelity $\mathcal{F}$. The result is
\begin{equation}
   \mathcal{N}_{123}(\mathcal{F})= \sqrt{ ( \mathcal{N}_{123}^{\mathrm{W}}\mathcal{F})^2 + ({1-\mathcal{F}} )^2  }-  (1-\mathcal{F}).
\end{equation}
In principle one can also calculate the tripartite negativity according to the density matrix solution above, analytically as well. However, these expressions become cumbersome due to the needed basis transformation before evaluating the eigenvalues of the partial transpose. As an example we describe the procedure for the initial state $\ket{\downarrow \uparrow\uparrow}$ in the sector $S^z=1/2$, so that the density matrix in the eigenstate basis is $\rho_{ij}^{S^z}(0)=1/3$ for all $i,j$. The bipartite negativities can then be written in terms of the matrix elements $\rho^{S^z=1/2}_{P,ij}(t)$ of the Pauli basis density matrix within the sector $S^z = 1/2$:
\begin{widetext}
\begin{align}
    \mathcal{N}_{[1]23} &= \sqrt{(1-\mathrm{Tr}[\rho^{S^z=1/2}_P(t)])^2 + 4(|\rho^{S^z=1/2}_{P,12}(t)|^2+|\rho^{S^z=1/2}_{P,13}(t)|^2)} - (1-\mathrm{Tr}[\rho^{S^z=1/2}_P(t)]),\\
    \mathcal{N}_{1[2]3} &= \sqrt{(1-\mathrm{Tr}[\rho^{S^z=1/2}_P(t)])^2 + 4(|\rho^{S^z=1/2}_{P,12}(t)|^2+|\rho^{S^z=1/2}_{P,23}(t)|^2)} - (1-\mathrm{Tr}[\rho^{S^z=1/2}_P(t)]),\\
    \mathcal{N}_{12[3]} &= \sqrt{(1-\mathrm{Tr}[\rho^{S^z=1/2}_P(t)])^2 + 4(|\rho^{S^z=1/2}_{P,13}(t)|^2+|\rho^{S^z=1/2}_{P,23}(t)|^2)} - (1-\mathrm{Tr}[\rho^{S^z=1/2}_P(t)]).
\end{align}
\end{widetext}
The trace is invariant under the basis transformation and thus $\mathrm{Tr}[\rho^{S^z=1/2}_P(t)] = \sum_i e^{-\gamma_i t}/3$. However, for the off-diagonal matrix elements directly entering the bipartite negativities, we need to explicitly perform the basis transformation $\rho_P(t) = U^\dag \rho U$ with
\begin{equation}
    U = \frac{1}{\sqrt{3}}
    \begin{pmatrix}
        1 & 1 & 1 \\
        1 & \eta & \eta^2 \\
        1 & \eta^2 & \eta
    \end{pmatrix},
\end{equation}
where $\eta = \exp(2\pi i/3)$. With this transformation one may explicitly write the density matrix elements as
\( \al{   \rho_{P,ij}^{S^z=1/2}(t) &= \frac{1}{3}\sum_{a,b=1}^3 \eta^{-(a-1)(i-1)}\rho_{ab}(t)\eta^{(b-1)(j-1)}\\ &= \frac{1}{9}S_{i-1}(t)S_{j-1}^*(t),} \)
where we defined  $S_k(t) = \sum_{a=1}^3e^{-\gamma_a t/2}e^{if_a t}\eta^{-k(a-1)}.$
We are only interested in the square moduli of these density matrix elements which can be very concisely written using the introduced quantities $S_k$ as
\begin{equation}
\al{    \rho_{P,12}^{S^z=1/2}(t) &= |S_0(t)|^2|S_1(t)|^2/81,\\
    \rho_{P,13}^{S^z=1/2}(t) &= |S_0(t)|^2|S_2(t)|^2/81,\\
    \rho_{P,12}^{S^z=1/2}(t) &= |S_1(t)|^2|S_2(t)|^2/81.}
\end{equation}
With the introduction of these new quantities, the bipartite negativities can be written as 
\begin{widetext}
\begin{align}
    \mathcal{N}_{[1]23} &= \sqrt{\left(1- \frac{1}{3}\sum_i e^{-\gamma_i t}\right)^2 + \frac{4}{81} |S_0|^2\big(|S_1|^2+|S_2|^2\big)} - \left( 1- \frac{1}{3} \sum_i e^{-\gamma_i t} \right),\\
    \mathcal{N}_{1[2]3} &= \sqrt{\left(1-\frac{1}{3} \sum_i e^{-\gamma_i t}\right)^2 + \frac{4}{81}|S_1|^2\big(|S_0|^2+|S_2|^2\big)} - \left( 1- \frac{1}{3} \sum_i e^{-\gamma_i t} \right),\\
    \mathcal{N}_{12[3]} &= \sqrt{\left(1- \frac{1}{3} \sum_i e^{-\gamma_i t}\right)^2 + \frac{4}{81}|S_2|^2\big(|S_0|^2+|S_1|^2\big)} - \left( 1- \frac{1}{3} \sum_i e^{-\gamma_i t} \right).
\end{align}
\end{widetext}
Generally, the values for $|S_k|^2$ are different for different $k$, and these bipartite negativities may therefore be different.

\section{Robustness of Entanglement Dynamics to Imperfections} \label{sec:non-homogeneous}
\begin{figure*}[!htbp]
\begin{subfigure}[t]{0.45\textwidth}
    \begin{overpic}[width=\linewidth]{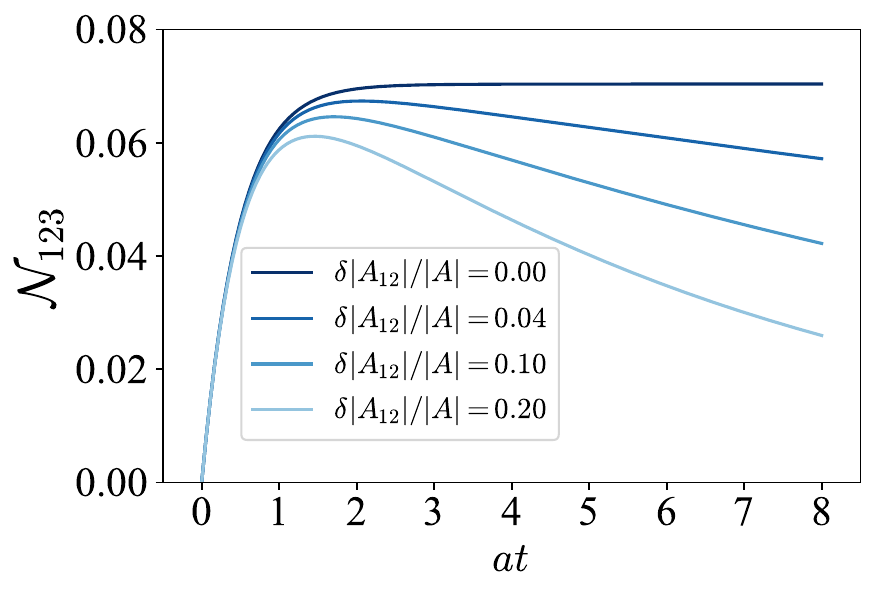}
        \put(0,68){(a)}
    \end{overpic}
\end{subfigure}
\begin{subfigure}[t]{0.45\textwidth}
\begin{overpic}[width=\linewidth]{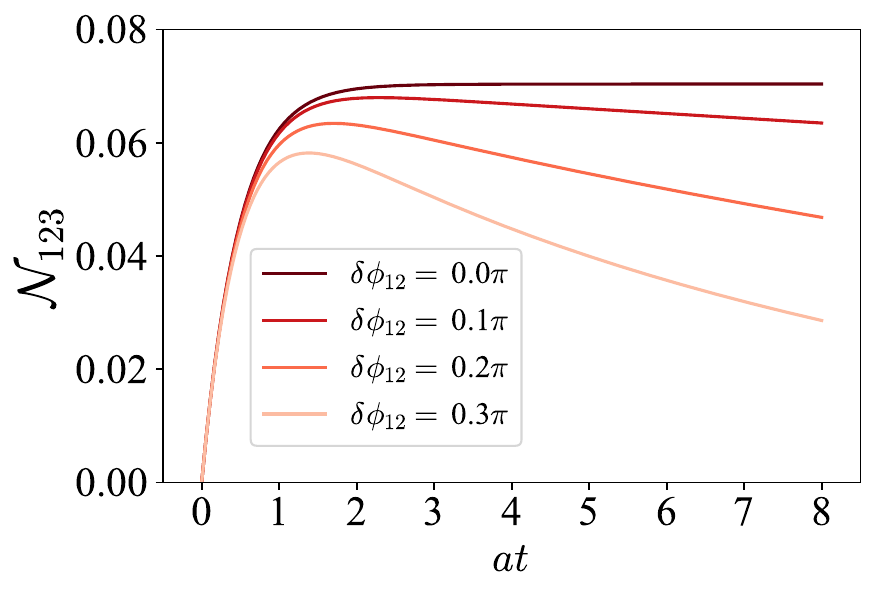}
        \put(0,68){(b)}
    \end{overpic}
\end{subfigure}
\caption{Entanglement dynamics for non-homogeneous non-local correlated noise for (a) varying correlated noise amplitude $|\delta A_{12}|$ and (b) phase $\delta \phi_{12}$. Initialization in a spin product state in the sector $S^z=1/2$, parameters such that $\gamma_0=0$ in the symmetric case, i.e. $\delta A_{12}=0, \delta \phi_{12}=0$. Parameters: $|A|/a=0.5$, $\phi=\pi$, $\mathcal{J}/a=0$. }
\label{figs1:two_figs}
\end{figure*}

In this appendix, we go beyond the ideal homogeneous limit discussed in Appendices \ref{sec:master_eq}-\ref{sec:tripartite_negativity} and investigate how various imperfections and control fluctuations affect the generation of tripartite entanglement. We focus on deviations that are relevant for realistic devices, including non-uniform correlated dissipation, asymmetries in the induced coherent interaction, additional dephasing channels, qubit-frequency detuning, and noise in the coherent drive. Throughout, we monitor the entanglement dynamics using the genuine tripartite negativity $\mathcal{N}_{123}$ defined in Appendix~\ref{sec:tripartite_negativity}.

\begin{figure*}[!htbp]
    \captionsetup[subfigure]{labelformat=empty}
    \centering
    \begin{subfigure}[t]{0.32\textwidth}
        \begin{overpic}[width=\linewidth]{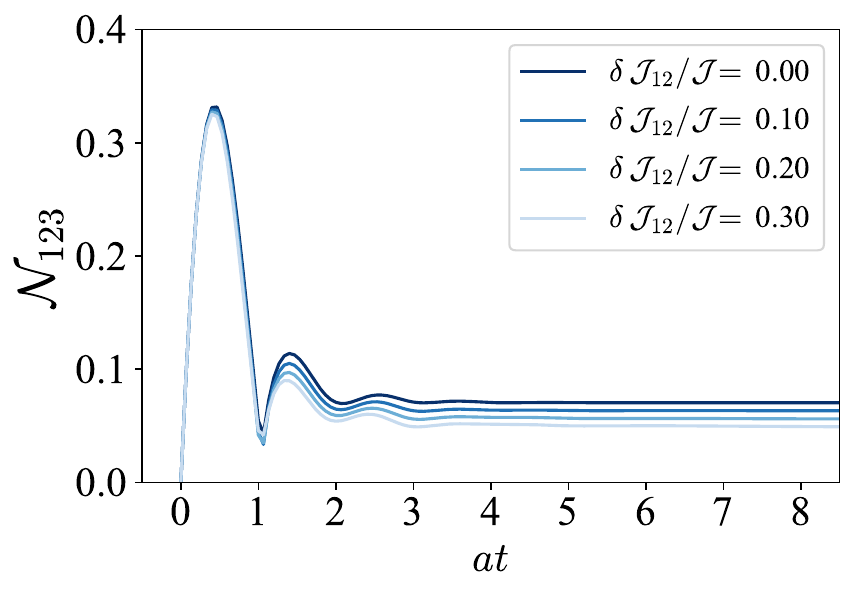}
            \put(0,68){(a)}
        \end{overpic}
        \caption[]{}
        \label{fig:delta_J12_psi0}
    \end{subfigure}
    \hfill
    \begin{subfigure}[t]{0.32\textwidth}
        \begin{overpic}[width=\linewidth]{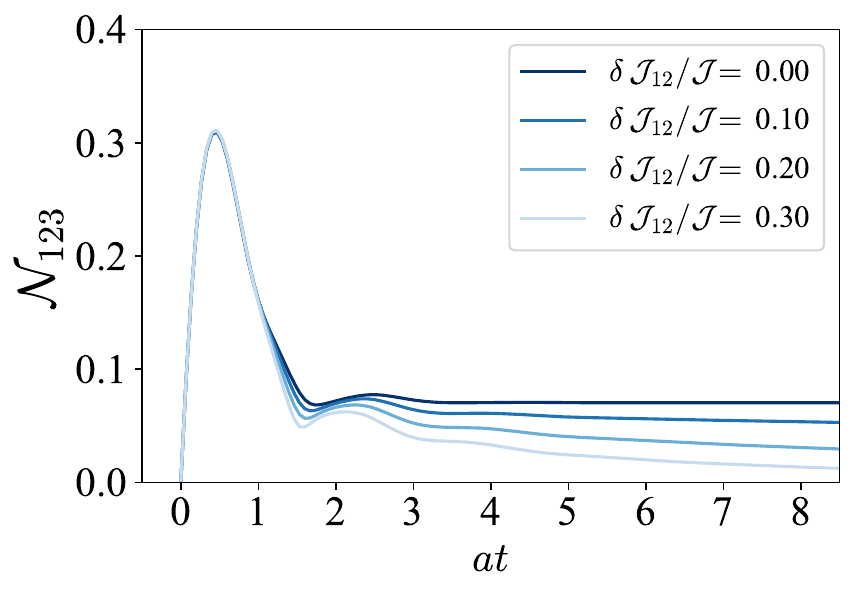}
            \put(0,68){(b)}
        \end{overpic}
        \caption[]{}
        \label{fig:delta_J12_psi03}
    \end{subfigure}
    \hfill
    \begin{subfigure}[t]{0.32\textwidth}
        \begin{overpic}[width=\linewidth]{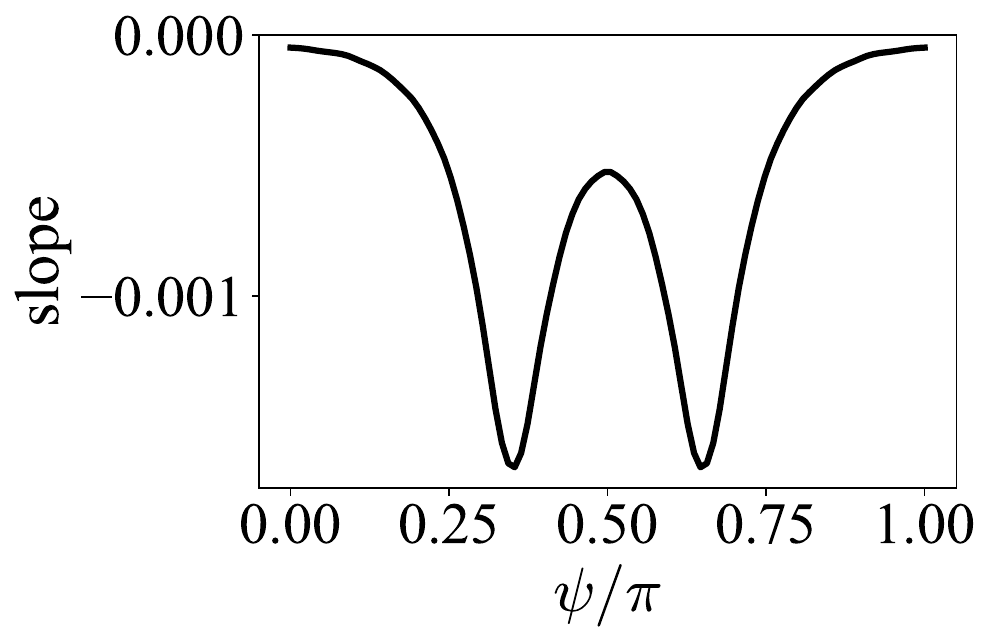}
            \put(0,68){(c)}
        \end{overpic}
        \caption[]{}
        \label{fig:delta_slope_psi}
    \end{subfigure}
    \caption{Dynamics of the tripartite negativity for different values of $\psi$ under an asymmetry in the effective coherent interaction described by Hamiltonian 
    $H' = \delta\mathcal{J}_{12}\Bigl(e^{i\psi}\sigma_1^+\sigma_2^- + e^{-i\psi}\sigma_1^-\sigma_2^+\Bigr).$
    (a) Time evolution for $\psi=0$. (b) Time evolution for $\psi = \pi/3$. (c) Decay rate (i.e. slope) of the tripartite negativity at time $at=12$ as function of $\psi$. Parameters: $|A|/a = 0.5, \mathcal{J}/a = 1, \phi = \pi$.}
\end{figure*}

Firstly, we explore the impact of having non-uniform correlated noise. To simulate this, we consider the case where the non-local dissipation on the link between qubits 1 and 2 differs from the remaining links. We therefore consider the noise correlation matrix
\begin{equation}
    \gamma = 
    \begin{pmatrix}
        a & A - \delta A_{12} & A^* \\
        A^* - \delta A_{12}^* & a & A \\
        A & A^* & a 
    \end{pmatrix},
\end{equation}
which means that the correlated noise associated with qubits 1 and 2 is slightly different from the noise affecting the other pairs. Because this modifies the matrix $\gamma$, the corresponding jump operators in the master equation also change. This directly influences the entanglement dynamics of the system.
Solving the master equation numerically, we find the multipartite entanglement dynamics in Fig.~\ref{figs1:two_figs}. Initially, at short times, the entanglement dynamics remains nearly unchanged. As time increases, the differences become more pronounced because the new jump operators allow the system to decay into the non-entangled ground state. Nonetheless, if the deviation $\delta A_{12}$ is small compared to $|A|$, the overall entanglement behavior is weakly affected.  



\begin{figure*}[!tb]
    \centering
    \begin{minipage}[t]{0.45\textwidth}
         \centering
         \includegraphics[width=\textwidth]{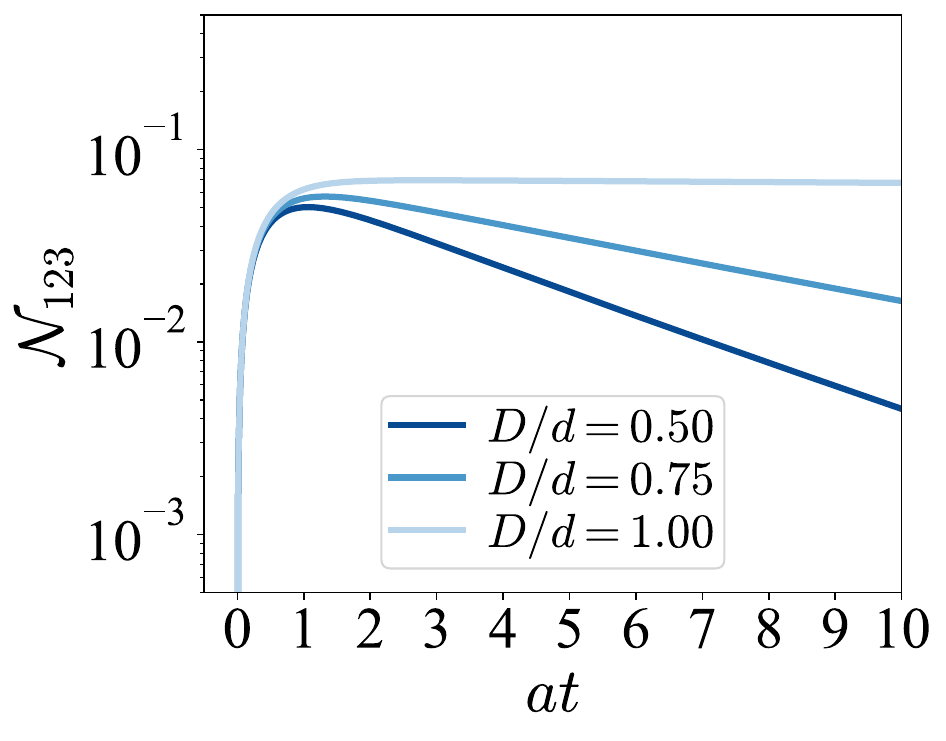} 
         \caption{\reapp{Tripartite negativity $\mathcal{N}_{123}$ as a function of time for finite local dephasing $d$ and varying collective dephasing strength $D$.  One observes that adding collective dephasing suppresses the decay of tripartite entanglement compared to purely local dephasing. Parameters: $|A|/a = 0.5$, $\mathcal{J}/a = 1$, $\phi = \pi, \varphi=0$.}}
         \label{fig:dephasing1}
     \end{minipage}
     \hfill 
     \begin{minipage}[t]{0.45\textwidth}
         \centering
         \includegraphics[width=\textwidth]{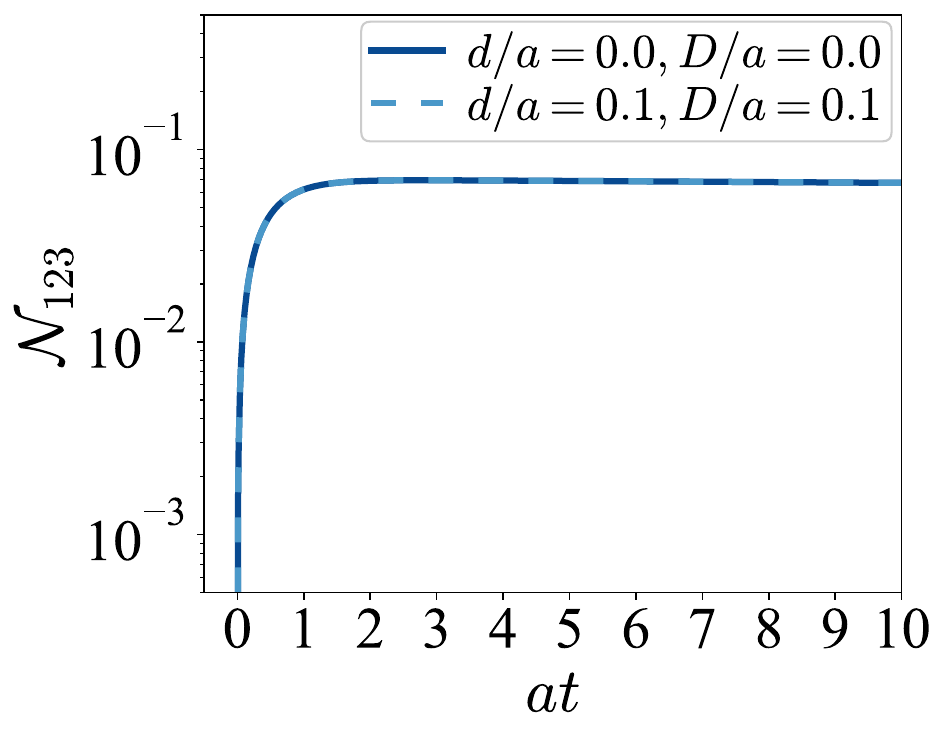} 
         \caption{\reapp{Tripartite negativity $\mathcal{N}_{123}$ as a function of time for dynamics without dephasing and for finite local and collective dephasing of equal strength.  Parameters: $|A|/a = 0.5$, $\mathcal{J}/a = 1$, $\phi = \pi, \varphi=0$. }}
         \label{fig:dephasing2}
     \end{minipage}

\end{figure*}

A non-homogeneous environment can also induce asymmetric coherent interactions in the effective Hamiltonian, such that $H_{\mathrm{eff}}\rightarrow H = H_{\mathrm{eff}} + H' $ with
\begin{equation}
    H' = \delta \mathcal{J}_{12}\big( e^{i\psi} \sigma^+_1 \sigma^-_{2} + e^{-i\psi} \sigma^-_1 \sigma^+_{2} \big).
\end{equation}
As discussed in the main text, the effective Hamiltonian $H_{\mathrm{eff}}$ has the $W$ states as eigenstates and therefore does not affect the entanglement dynamics. However, the $W$ states are not eigenstates of the perturbation $H'$. Instead,  for $\ket{\mathrm{W}} = \ket{S^z=1/2,\chi = 0}$, we obtain
\begin{equation}
    H'\ket{\mathrm{W}} = \frac{\delta \mathcal{J}_{12}}{\sqrt{3}}\big(e^{i\psi}\ket{\uparrow\downarrow} + e^{-i\psi}\ket{\downarrow\uparrow} \big)\ket{\uparrow} = \ket{\xi}.
\end{equation}
Hence, the perturbation maps the $W$ state with genuine tripartite entanglement to a state which is entangled only between qubit 1 and qubit 2. Thus, we expect that the perturbation will reduce the tripartite entanglement. As shown in Fig.~\ref{fig:delta_J12_psi0}, this is indeed the case, but the effect is not dramatic as long as the perturbation is small.
Also, the state $\ket{\xi}$ is not orthogonal to the $\ket{\mathrm{W}}$ which contrasts with the 2-qubit case. In the 2-qubit case the induced interaction takes you between parity sectors, which consequently leads to a fast decay to the energetic ground state which is not entangled. In the plots in Fig.~\ref{fig:delta_J12_psi0}, however, we observe that the tripartite entanglement saturates over time and the decay is negligible. The decay dynamics are fully governed by the jump operators $J_k$, which take you from one of the $W$ states to $\ket{\uparrow\uparrow\uparrow}$. Thus the relevant quantity to look at is the overlap between the state $\ket{\xi}$ and these $W$ states. We will denote the achiral $W$ state as $\ket{\mathrm{W}}$, which is the decoherence-free state, and the chiral $W$ states as $\ket{\chi = \pm 1}$. The overlaps are given by 
\begin{align}
    \bra{\mathrm{W}}\ket{\xi} &= \frac{2\delta \mathcal{J}_{12}}{3}\cos{\psi},\nonumber\\    \bra{\chi=\pm 1}\ket{\xi} &= \frac{2\delta \mathcal{J}_{12}}{3}\cos{(\psi \pm 2\pi/3)},\label{eq:overlap_psi}
\end{align}
such that the overlap between the decoherence free state and the state induced by the asymmetry is maximal at $\psi = 0$, leading to minimal decay to the energetic ground state $\ket{\uparrow\uparrow\uparrow}$. Therefore, with an increase in $\psi$, the decay is expected to increase, which is indeed observed in Fig.~\ref{fig:delta_J12_psi03}. To observe how this decay depends on $\psi$, we plot the slope at time $at=12$ of the tripartite negativity evolution in Fig.~\ref{fig:delta_slope_psi} as function of $\psi$. It may be observed that the decay is small, but non-zero, for $\psi=0$. This value for the slope varies very slowly with varying $\psi$, and therefore the decay dynamics shown in Fig.~\ref{fig:delta_J12_psi0} are robust to variations in the phase $\psi$ of the induced coherent interaction.

\begin{figure}[!htb]
 	\centering
	 \includegraphics[width=.82\linewidth]{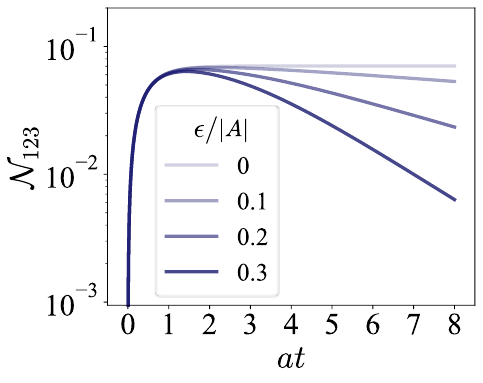}
         \caption{\reapp{Tripartite negativity plotted for finite detuning: it can be observed that the detunings can become rather significant with respect to the correlated noise before the dynamics change much. Parameters used: $|A|/a = 0.5, \mathcal{J}/a = 1, \phi = \pi$.}}
         \label{fig:detuning}
\end{figure}

\reapp{Next, we investigate the entanglement dynamics in the presence of pure dephasing channels arising from longitudinal couplings to the environment of the form $\sigma_\alpha^z\otimes E_\alpha$. For a homogeneous environment, the corresponding jump operators can be written as }
\begin{align}
	J_1 &=\sqrt{\frac{d+2|\mathcal{D}|\cos{(\varphi+2\pi/3)}}{3}}\nonumber\\ &\quad\times\left(\sigma_1^z e^{i2\pi/3} + \sigma_2^z e^{i4\pi/3} + \sigma_3^z \right), \\
	J_2 &=\sqrt{\frac{d+2|\mathcal{D}|\cos{(\varphi)}}{3}}\left(\sigma_1^z  + \sigma_2^z + \sigma_3^z \right), \\
	J_3 &=\sqrt{\frac{d+2|\mathcal{D}|\cos{(\varphi-2\pi/3)}}{3}}\nonumber\\ &\quad\times\left(\sigma_1^z e^{i4\pi/3} + \sigma_2^z e^{i2\pi/3} + \sigma_3^z \right) ,
\end{align}
\reapp{where $d$ and $\mathcal{D}$ are determined by local and non-local correlations of the environmental operators $E_\alpha$ at zero frequency, and $\varphi$ is the corresponding phase. We compare the dynamics for purely local dephasing ($\mathcal{D}=0$) with the case of significant collective dephasing $\mathcal{D}\sim d$. As shown in Fig.~\ref{fig:dephasing1}, local dephasing leads to a gradual decay of $\mathcal{N}_{123}$, while collective dephasing can strongly suppress this decay. In particular, when the correlated dephasing is the same as the local dephasing, the long-time tripartite entanglement is  unchanged compared with the purely dissipative case without dephasing, as observed in Fig.~\ref{fig:dephasing2}.}


\begin{figure*}[!htbp]
\captionsetup[subfigure]{labelformat=empty}
    \centering
    \begin{subfigure}[t]{0.37\textwidth}
        \begin{overpic}[width=\linewidth]{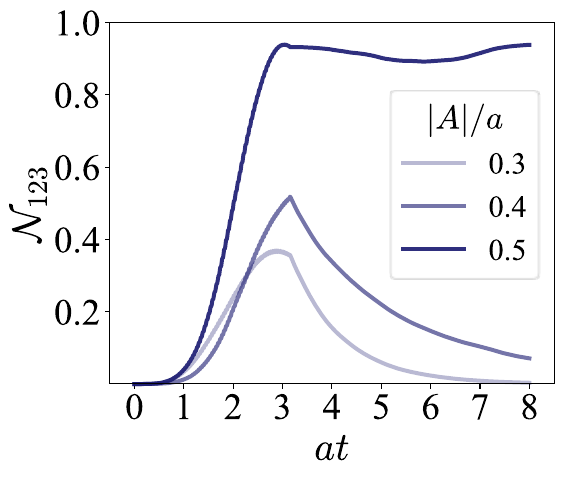}
            \put(0,80){(a)}
        \end{overpic}
        \caption[]{}
        \label{fig:amplitude_noise}
    \end{subfigure}
    \hspace{1.5cm}
    \begin{subfigure}[t]{0.37\textwidth}
        \begin{overpic}[width=\linewidth]{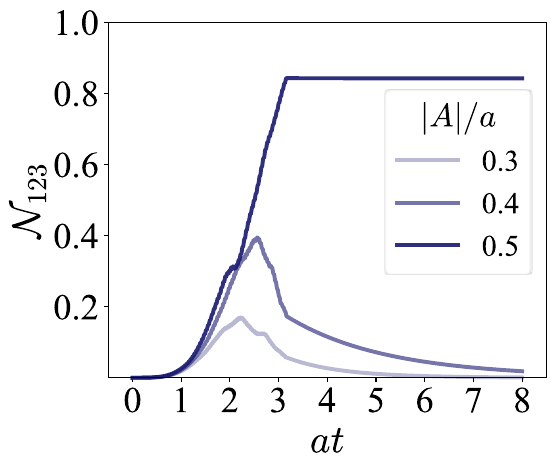}
            \put(0,80){(b)}
        \end{overpic}
        \caption[]{}
        \label{fig:frequency_noise}
    \end{subfigure}
    \caption{\reapp{Tripartite negativity for a noisy coherent drive. The curves correspond to varying correlated noise strength, with (a) noise parameters $\delta/a=0.1$ and $a\tau=1$, illustrating the impact of experimentally realistic amplitude fluctuations, {where $\tau$ is the correlation time of the noise $\xi(t)$}; and (b) with noise parameters $\delta/\mathcal{J}=0.05$ and $a\tau=1$, illustrating the impact of experimentally realistic frequency fluctuations on the build-up and decay of genuine tripartite entanglement. Parameters: $\mathcal{J}/a = 10, \phi = \pi$.}}
    \label{fig:coherent_drive_noise}
\end{figure*}

\reapp{We now relax the assumption of perfectly identical qubits and study the effect of detuning one of the qubit splittings. For concreteness, we add a small Zeeman shift on qubit 1, and consider the Hamiltonian
\begin{equation}
	H(\epsilon) = H_0 + \epsilon\sigma_1^z,
\end{equation}
where $H_0$ is the symmetric Hamiltonian used in Appendices \ref{sec:master_eq}-\ref{sec:tripartite_negativity}. The detuning breaks permutation symmetry and tends to decouple qubit 1 from the collective modes that form the W state. The results presented in Fig.~\ref{fig:detuning} show the evolution of tripartite entanglement for different values of $\epsilon/|A|$. We find that as long as the detuning is small enough compared to the correlated noise strength, the tripartite entanglement dynamics is only weakly modified. For larger detunings, the maximum $\mathcal{N}_{123}$ decreases and the long-time entanglement decays faster, reflecting a gradual crossover to effectively independent qubits.}

\begin{figure}[!htb]
 	\centering
	 \includegraphics[width=.82\linewidth]{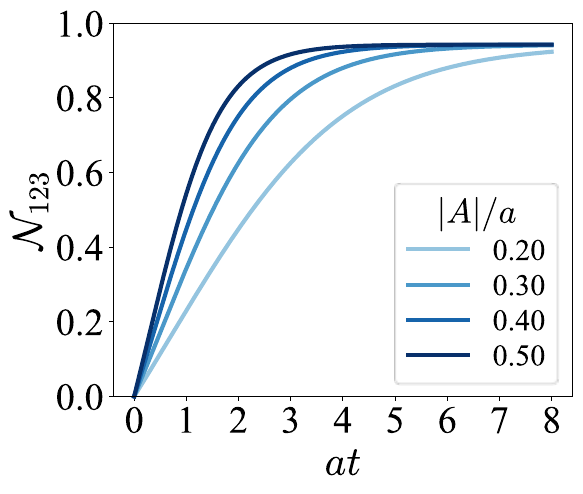}
         \caption{\reapp{Tripartite negativity with post-selection $\alpha=1$ for varying correlated noise strength: While the same maximum tripartite entanglement is ultimately reached for all $|A|$, smaller $|A|$ leads to a slower convergence towards this asymptotic state.  Parameters: $\mathcal{J}/a = 1, \phi = \pi$.}}
         \label{fig:post-selection_vary_A}
\end{figure}

\reapp{Moreover, we consider imperfections in the coherent-drive protocol used to boost the W-state fidelity. In realistic devices, both the amplitude and the frequency of the drive suffer from noise. To model amplitude noise, we replace the ideal drive amplitude $|C|/a=1$ by a stochastic amplitude,
\begin{equation}
	|C(t)|/a = 1 + \delta \xi(t),
\end{equation}
where $\xi(t)$ is a zero-mean Gaussian process with correlations $\langle \xi(t)\xi(s)\rangle = \exp{[-(t-s)^2/2\tau^2]}$. For the frequency noise, we let the detuning fluctuate as $\Delta+2\mathcal{J}\to \Delta+2\mathcal{J}+\xi(t)$ with the same statistics. Representative results for a moderate noise levels are shown in Figs.~\ref{fig:amplitude_noise} (amplitude noise) and \ref{fig:frequency_noise} (frequency noise). We observe that the maximal tripartite negativity is reduced but the qualitative build-up of entanglement remains intact. This indicates that the coherent-drive scheme is robust against realistic fluctuations, provided that the noise level does not exceed a few percent of the characteristic drive scale.}

\begin{figure}[!htb]
 	\centering
	 \includegraphics[width=.82\linewidth]{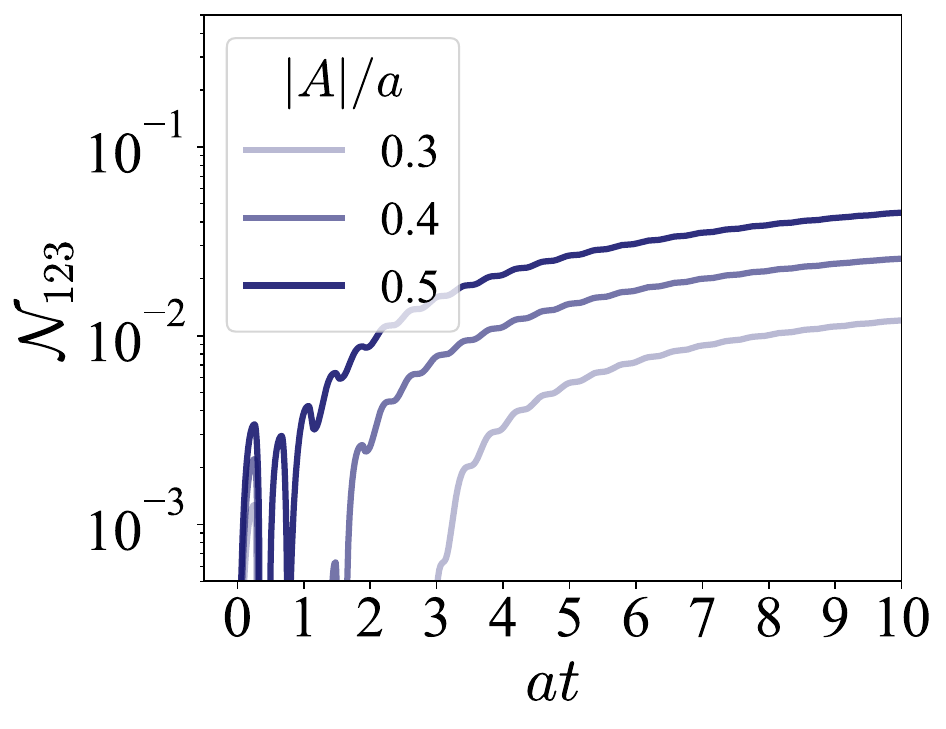}
         \caption{\reapp{Tripartite negativity under non-Markovian dissipation: The time-dependent dissipation rates lead to deviations from purely exponential Markovian behavior and give rise to oscillatory features in the entanglement dynamics, indicative of environmental memory and information backflow. Parameters used: $\mathcal{J}/a = 1, \phi = \pi$.}}
         \label{fig:non-markovian}
\end{figure}

\reapp{As a final robustness check, we vary the magnitude of the spatially correlated noise $|A|$ away from its optimal value $|A|=0.5a$ identified in the main text. For the post-selection protocol at zero temperature, we find that changing $|A|$ does not qualitatively change the dynamics. The same maximal tripartite entangled state is reached for all $|A|$, but smaller $|A|$ leads to a slower convergence to the asymptotic state (see Fig.~\ref{fig:post-selection_vary_A}).}

\begin{figure*}[!htbp]
\captionsetup[subfigure]{labelformat=empty}
    \centering
    \begin{subfigure}[t]{0.4\textwidth}
        \begin{overpic}[width=\linewidth]{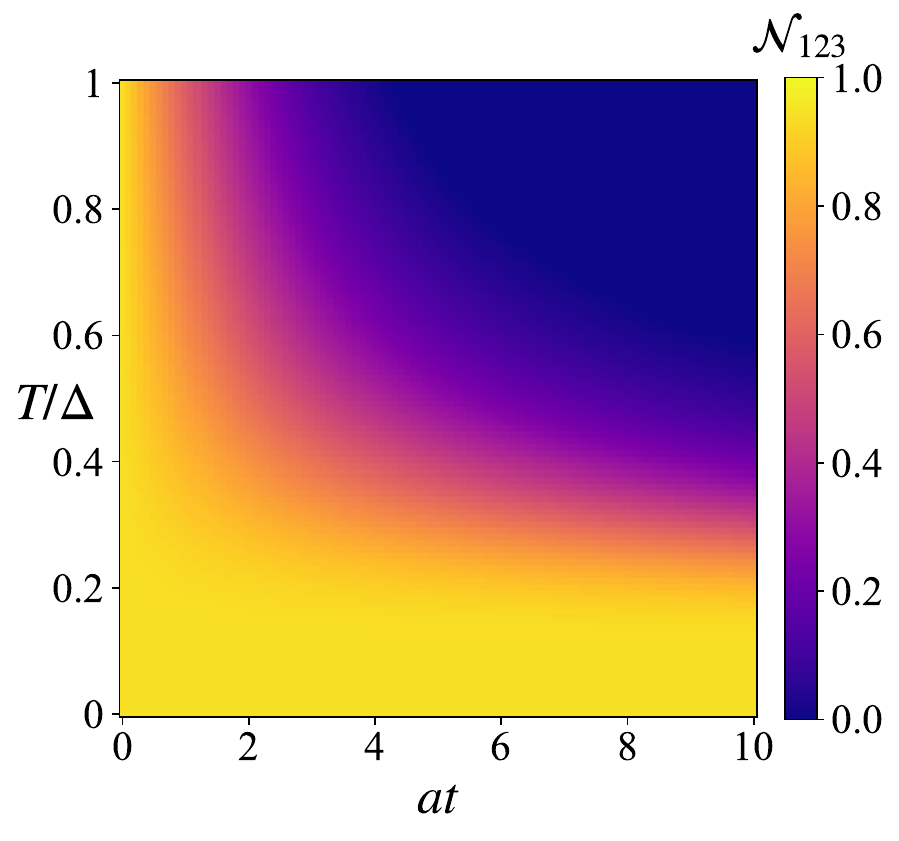}
            \put(0,90){(a)}
        \end{overpic}
        \caption[]{}
        \label{fig:fig1}
    \end{subfigure}
    \begin{subfigure}[t]{0.5\textwidth}
        \begin{overpic}[width=\linewidth]{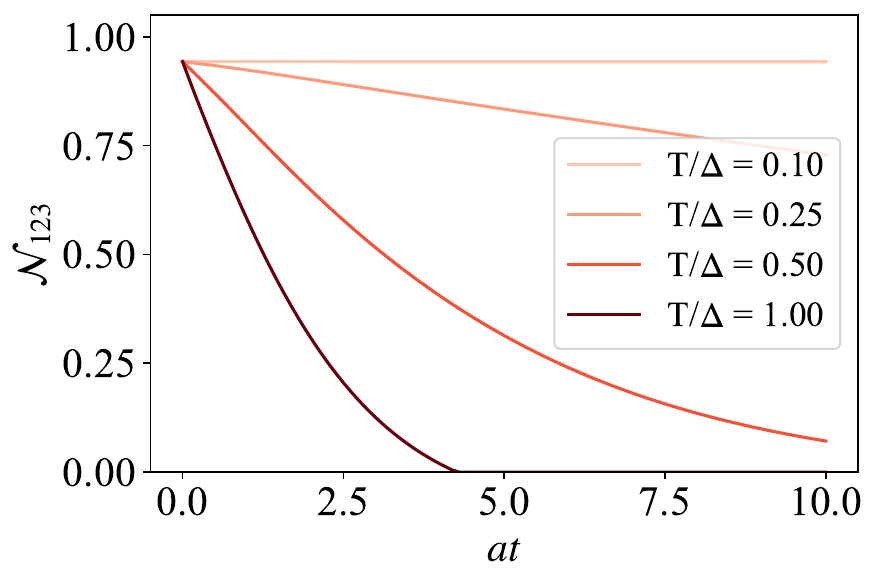}
            \put(0,72){(b)}
        \end{overpic}
        \caption[]{}
        \label{fig:fig2}
    \end{subfigure}
    \caption{Entanglement dynamics for varying temperature $T$ and initialization in $S^z=1/2$ sector in the eigenstate with $\chi = 0$, and choosing parameters such that $\gamma_0=0$. (a) Density plot of the entanglement as function of time and temperature. (b) Line cuts of the tripartite negativity at different temperatures. \hspace{\textwidth} Parameters: $|A|/a = 0.5, \mathcal{J}/a = 0, \phi = \pi$.}
    \label{fig:finiteTemperatureDynamics}
\end{figure*}

\reapp{So far, we have described the dynamics within the standard Markovian Lindblad framework, where the dissipation rates are time independent and the entanglement decays exponentially at long times. To assess the robustness of our entanglement-generation mechanism beyond this limit, we also considered a non-Markovian extension based on the time-convolutionless master equation~\cite{Zou2024SpatiallyQubits,gulacsi2023signatures}. In this approach, the environmental memory is encoded in explicitly time-dependent dissipation rates, leading to deviations from purely exponential decay.}

\reapp{Concretely, we implemented Eq.~(40) of Ref.~\cite{Zou2024SpatiallyQubits} for a $1/f$ noise spectrum, characterized by the parameters $\hbar/\sigma=10\,\mathrm{ns}, \omega_l/2\pi=1\,\mathrm{MHz},$ and $\Omega/2\pi=1\,\mathrm{GHz}$. The resulting tripartite negativity $\mathcal{N}_{123}(t)$ is shown in Fig.~\ref{fig:non-markovian}, where one clearly observes oscillatory features induced by the time-dependent rates. These oscillations are signatures of environmental memory and information backflow from the bath to the system. While the overall entanglement generation mechanism remains qualitatively the same as in the Markovian case, non-Markovian effects can transiently enhance or suppress the entanglement compared to a purely exponential decay which is the signature of Markovian dynamics.}

\begin{figure}[!htb]
 	\centering
	 \includegraphics[width=\linewidth]{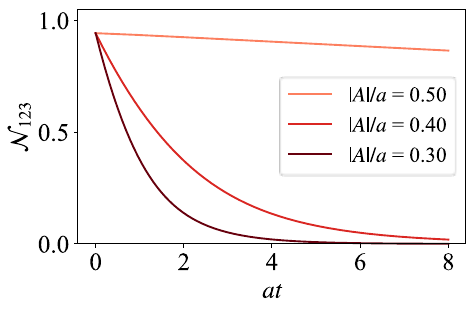}
         \caption{\reapp{Tripartite negativity at finite temperature ($T/\Delta=0.2$) for different correlated noise strength: In this finite-temperature regime, the dynamics are highly sensitive to $|A|$: for $|A|/a=0.5$ the entanglement decays only slowly, whereas for smaller $|A|$ it shows an almost exponential decay, strongly suppressing the long-time tripartite entanglement. {At $T=0$, the $W$ state ($|A|/a = 0.5$)  becomes a true dark state and $\mathcal{N}_{123}$ remains constant, while
    for $|A|/a < 0.5$ it decays exponentially with lifetime
    $\sim 1/(a - 2|A|)$.} Parameters used: $\mathcal{J}/a = 1, \phi = \pi$.}}
         \label{fig:finite_T_vary_A}
\end{figure}

\section{Entanglement Dynamics at Finite Temperature} \label{sec:finite_temp}

In this appendix, we briefly discuss the entanglement dynamics at finite temperature $T$. We assume that the system is prepared in the highly entangled achiral $W$ state $|S^z = 1/2, \chi = 0\rangle$ at $t=0$. We further choose the parameters such that the amplitude $\gamma_0 = 0$. At zero temperatures,the system therefore remains in the initial state. At non-zero temperature, however, thermal excitations can trigger processes to the sector $S^z = -1/2$, followed by decay towards the energetic ground state  $\ket{\uparrow\uparrow\uparrow}$ of the system. As a consequence, the finite temperature destroys the entanglement. The resulting entanglement dynamics is shown in Fig.~\ref{fig:finiteTemperatureDynamics} for different temperatures. For sufficiently low temperature, large multipartite entanglement can persist over a significant period of time.

\begin{figure}[!htb]
 	\centering
	 \includegraphics[width=.82\linewidth]{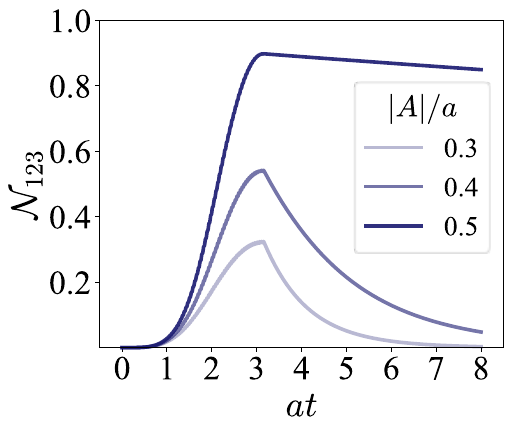}
         \caption{\reapp{Tripartite negativity under coherent drive at finite temperature $T/\Delta=0.2$: Compared to the zero-temperature case, finite temperature reduces the maximal achievable entanglement negativity and fidelity but the entanglement generation remains robust and the qualitative dynamics are preserved.} {The corresponding zero-temperature dynamics under the same drive protocol is shown in Fig.~\ref{fig:3}(d).} Parameters used: $\mathcal{J}/a = 10, \phi = \pi$.  }
         \label{fig:drive_finite_T}
\end{figure}

\reapp{To investigate how finite temperature affects the dissipative entanglement generation protocol for non-local dissipation deviating from $|A|/a=0.5$, we also simulated the dynamics at $T/\Delta=0.2$ for different values of the correlated noise strength $|A|$. The resulting entanglement dynamics are shown in Fig~\ref{fig:finite_T_vary_A}. When $|A|/a=0.5$ the entanglement decays only slowly in time. Detuning $|A|$ from this optimal value (here by 20\% and 40\%) leads to a noticeably faster decay, approaching an almost exponential suppression of the long-time tripartite entanglement. This behavior demonstrates that, in the finite-temperature regime, maintaining $|A|$ close to its optimal value is important to ensure long-lived multipartite entanglement.}

\reapp{We also studied the effect of finite temperature on the coherent-drive protocol that enhances the W-state fidelity. Figure~\ref{fig:drive_finite_T} shows $\mathcal{N}_{123}(t)$ for a temperature of $T/\Delta=0.2$ using the same drive parameters as in Appendix~\ref{sec:non-homogeneous}. Compared to the zero-temperature case, finite temperature reduces the maximal achievable entanglement negativity and fidelity, since thermal excitations populate additional states that do not participate in the driven W-state manifold. Nevertheless, the entanglement generation remains robust and the qualitative dynamics are preserved: the coherent drive still produces a pronounced entanglement peak and a long-lived plateau. For the post-selection approach, we expect the influence of finite temperature to be weaker because post-selection effectively filters out trajectories with thermal excitations; in that case the maximum entanglement can still be reached, although the approach to the asymptotic state is slower. Overall, both the dissipative and drive-assisted schemes remain effective in the experimentally relevant temperature range $T/\Delta\sim 0.1-0.5$.}

\FloatBarrier
%

\end{document}